\documentclass{article}%
\usepackage{amsfonts}
\usepackage{amsmath}
\usepackage{amssymb}
\usepackage{graphicx}%
\providecommand{\U}[1]{\protect\rule{.1in}{.1in}}
\newtheorem{theorem}{Theorem}

\newtheorem{definition}[theorem]{Definition}

\begin{document}

\title{A critical look at 50 years particle theory from the perspective of the
crossing property\\{\small Dedicated to Ivan Todorov on the occasion of his 75th birthday}\\{\small to be published} {\small in Foundations of Physics} }
\author{Bert Schroer\\CBPF, Rua Dr. Xavier Sigaud 150 \\22290-180 Rio de Janeiro, Brazil\\and Institut fuer Theoretische Physik der FU Berlin, Germany}
\date{December 2009 }
\maketitle

\begin{abstract}
The crossing property is perhaps the most subtle aspect of the particle-field
relation. Although it is not difficult to state its content in terms of
certain analytic properties relating different matrixelements of the S-matrix
or formfactors, its relation to the localization- and positive energy spectral
principles requires a level of insight into the inner workings of QFT which
goes beyond anything which can be found in typical textbooks on QFT. This
paper presents a recent account based on new ideas derived from "modular
localization" including a mathematic appendix on this subject. Its main novel
achievement is the proof of the crossing property of formfactors from a
two-algebra generalization of the KMS condition.

The main content of this article is the presentation of the derailments of
particle theory during more than 4 decades: the S-matrix bootstrap, the dual
model and its string theoretic extension. Rather than being related to
crossing, string theory is the (only known) realization of a dynamic infinite
component one-particle wave function space and its associated infinite
component field. Here "dynamic" means that, unlike a mere collection of
infinitely many irreducible unitary Poincar\'{e} group representation or free
fields, the formalism contains also operators which communicate between the
different irreducible Poincar\'{e} represenations (the levels of the "infinite
tower") and set the mass/spin spectrum. Wheras in pre-string times there were
unsuccessful attempts to achieve this in analogy to the O(4,2) hydrogen
spectrum by the use of higher noncompact groups, the superstring in d=9+1,
which uses instead (bosonic/fermionic) oscillators obtained from
multicomponent chiral currents is the only known unitary positive energy
solution of the dynamical infinite component pointlike localized field project.

When the first version of this paper was submitted to hep-th it was
immediately removed by the moderator and placed on phys. gen without any
possibility to cross list, even though its content is foundational QFT. With
the intervention of another member of the advisory comitee at least the
cross-listing seems now to be possible.

\end{abstract}

\section{The increasing gap between foundational work and particle theory}

There has always existed a tendency to romanticize the past when criticizing
the present. But the importance of interpretational and philosophical ideas
for the development of quantum theory (QT) in the first three decades of
particle theory, starting in quantum mechanics (QM) and escorting the
beginnings of quantum field theory (QFT), as compared to their superficial
role or absence in the ongoing particle theory is hard to be overlooked. Most
of the foundational concepts in relativistic QT can be traced back to
developments before 1980. One can hardly think of any other branch of physics
in which the correct interpretation of observational results was that much
dependent on the outcome of a delicate balance between speculative innovations
being followed by critical foundational work in which questions of conceptual
aspects and philosophal consistency were the main driving force.

The strength of this connection between descriptive and conceptual aspects in
the beginning of quantum physics was a result of the protagonist's (Bohr,
Heisenberg, Schroedinger,..) intense interests in conceptual and philosophical
questions of of quantum theory. Almost the entire arsenal of foundational
concepts, including those iconized Gedankenexperiments as
\textit{Schroedinger's cat} and \textit{Fermi's two-atom experiment} in QED
(arguing that $c$ as the classical velocity of light remains the maximal speed
after the quantization of electrodynamics), were introduced in order to
highlight the philosophical consequences of their discoveries and to
facilitate a critical engagement with the new theory for others.

But this does not mean that all this impressive grand design was an inevitable
outcome of the innovative potential of the protagonists. Even the greatest
intellectual brilliance is no insurance for finding the "diretissima" for
scientific progress; already a slight change in the chronological ordering of
important discoveries could have led to a time-consuming detour or a blind allay.

Just imagine that Feynman's path integral would have entered the scene before
matrix mechanics, Schr\"{o}dinger QM and transformation theory; as a result of
the conceptual proximity of an integral over classical orbits with the
Bohr-Sommerfeld framework of the largely quasi-classical old quantum theory
there is hardly anything more natural than to contemplate such a direct
connection. The resulting formalism would have unified all the quasi-classical
results of the old quantum theory and lifted it to a new level. It would have
streamlined most previous calculations and presented an elegant way how to do
computations around quantum oscillators, but it would have missed the
important dichotomy between observables and states. Even worse, the
elaboration of the Hilbert space formalism and operators acting in it, as well
as all the understanding of those important integrable systems as the hydrogen
atom (which even with all the present hindsight about path integrals remaines
a nontrivial endeavour) without whose operator presentation a course on QM is
unthinkable, all these important contributions would have appeared much later
and in a very different (and probably more involved) form.

Fortunately this was not the way things unfolded; by the time Feynman proposed
his path representation in the setting of QM, the conceptual level of operator
QT was mature enough to resist the temptation of a fallacious short-sighted
interpretation of this elegant, but often conceptual and computational unsafe
formalism. In this way many years of confusion in quantum physics were avoided
and the path integral could be explored for those purposes for which it is
powerful, namely quasiclassical approximations, keeping track of combinatorial
aspects of renormalized perturbation theory\footnote{Despite the fact that it
is not valid in interacting QFT, its is an ideal intuitive starting point and
with some hinsight about the nature of renormalization it carries its user
right into the renormalized perturbation formalism. Even though the result
does not satisfy the path representation from which everything started, it was
a valuable guide to arrive at the correct result.} and for presenting a
flexible metaphoric top soil on which innovative ideas can sprout and specific
computational problems be formulated.

Many results based on operator formalism on the other hand are either out of
reach of the path integral, or can only be obtained by imposing rules which do
not follow from its measure theoretic foundation and are therefore less
trustworthy than direct operator methods. The conceptual-mathematical control
is limited to QM and certain (superrenormalizable) models in low dimensional
QFTs, but this does not diminish its value as an intuitive guide and a social
cohesion-creating construct in discussions among particle physicists with
different backgrounds.

Taking into account that progress in particle physics is not only the result
of the intellectual capacity and the originality of the involved actors, but
also requires an element of good fortune in taking the right turns at the
right time on important cross roads, there is ample reason for considering the
first three decades of particle physics in retrospect as the "good old days".
The aim of this essay is to shed light on later developments, when innovation,
critical analysis and luck began to drift apart. The best way to do this is to
revisit the chain of events which started from the S-matrix bootstrap approach
and culminated in string theory.

It is not difficult to localize the point of no return, from where the present
less fortunate direction in particle theory research took its beginning, by
following the events in the aftermath of the enormous successful perturbative
renormalized quantum electrodynamics (QED). The emerging difficulties to treat
the nuclear interactions with the same methods led to a revival of S-matrix
based ideas. This time the connection between relativistic local fields and
asymptotic in/out particles was better understood than in Heisenberg's
ill-fated first attempt \cite{Heisen} a decade before the S-matrix bootstrap.

Instead of investigating a concrete hadronic model, for which there existed at
that time no computational framework, the most reasonable approach was to look
for some experimentally accessible consequences of the general causal locality
principles underlying QFT. This led to the derivation of a form of the
Kramers-Kronig dispersion relations known from optics, but now ingeniously
adapted to particle physics. The derivation of these relations from first
principles and their subsequent experimental verification in high energy
collisions was the main aim in which many of the best brains of the 50s participated.

According to the best of my knowledge this was the only topic in post QED
particle theory which can be characterized by the words "mission
accomplished"; several years of dedicated work led to the solution of the
problem, so that one could move on to other problems in an upbeat spirit
without being obliged to revisit the problems in order to patch up conceptual
holes left behind.

It was in the wake of dispersion theory that the notion of the crossing
property appeared; first as a property in Feynman graph perturbation theory
and soon afterwards as a consequence of the same principles which already led
to the dispersion relation. Bros Epstein Glaser and Martin \textbf{\cite{BEG}%
\cite{E-G-M}} succeeded to proof the validity of crossing property by showing
that the two particle elastic scattering amplitude is analytically connected
to its crossed\footnote{An incoming particle changes its position with an
outgoing one and, as required by charge conservation, both particles become
anti-particles.} version. The analytic connection between these processes
establishes the existence of a "masterfunction" which analytically links all
these different processes. Its existence in turn suggested that the asymptotic
high energy behavior of the different processes may not be independent, an
idea which was confirmed in \cite{Martin}. There exist also proofs of
"asymptotic crossing" for $2\rightarrow3~$scattering and indications about how
to generalize this to $2\rightarrow n$ scattering \cite{Br}. Some comments on
the ideas used in this derivation can be found in the next section.

Since causal localization is the only foundational property which
distinguishes QFT within quantum theory (for this reason often referred to as
LQP i.e. local quantum physics \cite{Haag}), the fact that the wealth of
different models with their distinct physical manifestations are in some way
related to different realizations of causal localization is to be expected.
What is however highly nontrivial is the chain of arguments and the richness
of additional concepts which are needed in order to establish this connection.

In the present work one of the oldest and most mysterious properties on the
border between particle and fields, namely the crossing property, is
generalized to formfactors and general scattering amplitudes. The modular
localization methods used in that derivation reveal that the conceptual
setting is a \textit{two-algebra generalization of the thermal KMS property
}(section 5). Although this KMS property is, as the Bros Epstein Glaser
arguments which are based on analytic completions of expectation values,
derived from locality and localization, the former is easier, furthergoing and
more physical.

Continuing the S-matrix history, in the subsequent revival of S-matrix theory
the newly discovered crossing played an essential role. It was the main
distinctive new feature with respect to Heisenberg's ill-fated prior S-matrix
proposal of the 40's. The S-matrix bootstrap program attracted the attention
of many particle theorists for almost a decade before it disappeared from
journal publications and conference topics\footnote{The fate which the
S-matrix bootstrap community in conference publications predicted for QFT was
"to fade away like a mortally wounded soldier on a battle field", but little
did they know that this would become its own fate shortly after.}. The
apparent reason was "physical anemia" i.e. its inability to produce any
credible calculation from its underlying principles. There was certainly
nothing wrong with its S-matrix principles of unitarity, Poincar\'{e}
invariance and crossing, except that the "maximal analyticity" postulate
resulted from a misunderstanding of the role of analyticity in physics since
different from the times of the S-matrix bootstrap, it not represent a
physical principle but rather results from one which unfortunately was not
clearly identified. The connection between the physical causal locality
principles and their analytic consequences are subtle and long winding, but
there is no way to sidestep these subtleties by turning the logic on its head.

What was however grossly misleading was the claim that the \textit{nuclear
democracy}\footnote{The quantum mechanical hierarchy between elementary and
bound particles cannot be maintained in QFT; the only hierarchy which is
consistent with interaction caused vacuum polarization clouds is that between
basic and fused superselected charges.} behind the bootstrap principles has at
most one solution (the possibility of having no solution was admitted) which
describes the entire world of strong interaction. Such sweeping
ultra-reductionist uniqueness claims arose occasionally in particle physics,
usually in connection with certain nonlinear structures to which it was
difficult to find any solution at all (e.g. the Schwinger-Dyson equation).
Reasonable formulations as QFT "defuse" such nonlinear structures as e.g. the
unitarity of the S-matrix by showing that they result from linear asymptotic
properties of fields.The belief in the uniqueness of the bootstrap mechanism
contained already germs of modern ideological thinking, which in more recent
times took the extreme form of a theory of everything (TOE).

Several years after the disappearance of the S-matrix bootstrap, the
principles which underlie the construction of so-called factorizing
two-dimensional models were discovered \cite{KTTW} which kick-started a still
ongoing stream of results about a family of new interesting soluble
models\footnote{See the most recent one \cite{Foer} and the references quoted
therein.}. These rich results came from the observation that factorizing
two-dimensional elastic S-matrices can indeed be classified and constructed by
the bootstrap principles of the meanwhile abandoned S-matrix bootstrap
approach. Factorization in conjunction with dispersion theoretic analyticity
led to meromorphy in terms of the rapidity variables which gave a precise
meaning and a physical interpretation to "maximal analyticity" at least in
this special case. The protagonists of the old bootstrap program never took
notice of these astonishing new observations about an interesting subset of
nontrivial QFTs which have become a theoretical laboratory to test ideas of
QFT \cite{Sch}\cite{Lech1}; in this way they spared themselves the
confrontation with their earlier premature apodictic statements on this matter.

The two-dimensional bootstrap project has infinitely many solutions and serves
as the starting point of a new infinitely large family of genuine nontrivial
two-dimensional QFTs. These constructions did not only show that the claimed
unicity was wishful imagination, but also revealed that the idea that all QFT
can be described in a Lagrangian setting was too narrow: the bootstrap
classification of all two-dimensional factorizing S-matrices had infinitely
many more solutions than those which can be described by pointlike Lagrangian
couplings between free fields.

There were many ad hoc concepts invented in the wake of the S-matrix
bootstrap, the most prominent (used in many later papers) was the Mandelstam
spectral representation \cite{Vech}. At that point the philosophy underlying
physical research had significantly changed as compared to the era of
dispersion relation\footnote{I recall warnings by K\"{a}ll\'{e}n, Lehmann,
Jost, Martin and others.}. For the latter it was essential to be a
\textit{rigorous} consequence of spectral representation (the
Jost-Lehmann-Dyson representations) which in turn were derived from the
locality and spectral principles of QFT. Without this strong connection with
the underlying principles, the experimental verification of dispersion
relations would have remained without much significance since they represented
a check of the locality principles of QFT and not of the validity of a
particular model.

The aim of the work of Mandelstam, as well as that of the later work of
Veneziano leading up to string theory, was very different from that of
dispersion relations; although it originated with a phenomenological
entitlement, it soon turned into a rather freewheeling attempt to explore an
imagined area beyond QFT with yet unknown principles. In other words these
attempts were excursions into the "blue yonder", but certainly not from a firm
platform of departure to which one could return in case of failure which in
particle theory is more common than success. As soon as the phenomenological
basis was lost as a result of new experiments which turned out to be
incompatible with the Regge trajectory phenomenology, the dual model and
string theory became free-floating mathematical ideas without any conceptual
basis to which they could safely return.

The main part of the paper will be concerned with a critical look at post
S-matrix bootstrap ideas as the phenomenological dual model and the closely
related string theory, which the protagonists of these models and others
thought of as particular implementations of the crossing property. Following
\cite{Mack} it will be shown that the dual model properties are identical to
the analytic properties of Mellin transforms of conformal correlation; they
have nothing in common with the correctly understood crossing property of
formfactors and scattering amplitudes which belong to a very different
conceptual setting. Since the \textit{crossing property is one of the most
subtle relations between particles and fields}, part of our task consists in
presenting an up to date account of a derivation of crossing from the
causality and covariance principles of QFT.

The full depth of the crisis in contemporary particle physics cannot be
perceived, and its causes cannot be understood, without a careful conceptual
and mathematical analysis based on a critical first hand historical knowledge.
Commemorative articles as \cite{Vech} are interesting and certainly contain a
lot of important background material, but one should not expect to find a
critical view in them.

If one asks a particle theorist of sufficient age to point at an important
difference between the scientific discourse in the old days and the one in
more recent decades, he will probably agree that, whereas the intellectual
potential has remained the same or even increased, there has been a remarkable
reduction of critical contributions and public controversies. The great
conceptual discourse of the early years of QT gave way to a new style in which
metaphorical arguments were allowed a more permanent position and the
appreciation of the pivotal role of\ the delicate equilibrium between
innovative speculations and their critical evaluation (which made particle
physics such a success story) was declining.

At the time of Pauli, Lehmann, K\"{a}ll\'{e}n, Feynman, Landau, Jost,
Schwinger and others it was the critical analysis of new ideas which kept
particle theory on a good track. Although controversies became sometimes
abrasive for the persons directly involved, particle physics profited from
them. Since the time of Jost's criticism \cite{Jost} of the S-matrix bootstrap
idea in the 60s, there has not appeared any profound critical article of essay
about the ideas leading from S-matrix theory to string theory\footnote{By this
I mean primarily an inner theoretical critical discourse clarifying the
conceptual position with respect to the principles underlying previous
successful theories.}. Less than ever was string theory itself subjected to
critical evaluation about its conceptual-mathematical structure, the critique
of its sociological and philosophical epiphenomena is not sufficient. Those
prestigious physicists, who in previous times would have considered it as
their privilege, if not their moral duty, to give a critical account, became
string theories fiercest defenders, if not to say its propagandists.

For a historical and foundational interested researcher with textbook
knowledge of QFT, the 40 year lasting dominance of this theory is surrounded
by a nearly impenetrable mathematical conceptual cordon which makes it
difficult to extract relevant foundational aspects. The present article can
not change a situation which has been going on for 40 years and in this way
became immunized against conceptual objections \footnote{In the words of
Feynman: "string theory has no arguments instead it uses excuses",}, but it
does present some unknown facts which may become useful in a not so far
future, when historians and philosophers finally become curious about what
really went on in particle physics for almost half a century and in particular
what happened to all those noisy promises of a TOE.

The content of the various sections is as follows. The next section explains
the formal aspects of the crossing property. It contains in addition to
mathematical facts also philosophical aspects. The third section presents the
dual resonance model and its derivatives and explains why the absence of a
critical evaluation of this interesting class of models which result from the
suitably normalized Mellin transformation of (any) conformal QFT (and have
nothing to do with properties of the S-matrix) prepared the ground which led
into the metaphoric landscape on which string theory subsequently flourished.
The fourth section shows that string theory is, despite its name, not about
objects which have a string-like spacetime localization; rather the objects of
string theory define a \textit{"dynamical" infinite component pointlike field
}\footnote{The "dynamical" has been added in order to distinguish the intended
meaning from the trivial case of an infinite direct sum of irreducible free
field representations. In addition to such an infinite mass/spin tower there
are also \textit{intertwiners between these representations} without which one
cannot generate a mass/spin spectrum.}; this section therefore constitutes the
core of the critical part of the presentation.

Section 5 and 6 present the modern view of the crossing property which to a
certain extent explains why it led to so many misunderstandings and metaphoric
ideas. Despite the highly mathematical level of these sections, the
presentation of the mathematical state of art on crossing is not the principle
motivation. But a critical exposition of ideas which historically emanated
from an incompletely or even incorrectly understood crossing property would
itself be incomplete without giving the modern viewpoint on this subtle
property. The conclusions present a resum\'{e} and additional critical remarks.

\section{The crossing property and the S-matrix bootstrap approach}

In contrast to QM where particles play the role of stable quanta which keep
their identity in the presence of interactions, QFT comes with a much more
fleeting particle concept. Even in theories without interactions, where
relativistic particles are synonymous with free fields, composite operators as
e.g. the important conserved currents exhibit the phenomenon of (finite)
vacuum polarization. This makes such an object rather singular (an
operator-valued distribution with no equal time restriction) and renders the
definition of a \textit{partial charge} corresponding to a finite volume a
delicate problem with the help of which Heisenberg \cite{Hei} discovered the
property of vacuum polarization at the beginning of QFT.

The full subtlety of this problem only became manifest in the presence of
interactions; this is the situation in which Furry and Oppenheimer \cite{F-O}
observed that even the basic Lagrangian fields, which without interactions
were linear in the particle creation/annihilation operators, cannot create
one-particle states without an admixed infinite cloud\footnote{In the sequel
"cloud" is intended to automatically imply an infinite number of particles.}
of particle/antiparticle pairs. Re-interpreted in a modern setting, this
observation permits the following generalization\textit{: in an interacting
QFT there exists no operator localized in a compact spacetime region which, if
applied to the vacuum, creates a one-particle state without an infinite vacuum
polarization cloud.} Or using recent terminology: a model which contains among
its operators a compactly localized PFG (vacuum-\textbf{p}%
olarization-\textbf{f}ree \textbf{g}enerator) is generated by a free field
\cite{BBS}\footnote{The theorem is the algebraic version of the Jost-Schroer
theorem, see \cite{STW}. The latter shows that the existence of a local
covariant field which acts on the vacuum as PFG implies that it is a free
field whereas the former replaces the pointlike covariance with the
affiliation to a compact localized algebra.}. The "shape" of the locally
generated vacuum polarization cloud depends on the kind of interaction, but
its infinite particle content is a characteristic property shared by all
interacting theories; a finite number of particle-antiparticle polarization
pairs created by "banging" on the vacuum with a local (composite) operator can
only happen in a free theory. The sharpness of the localization boundary
(horizon) accounts for the unboundedness of the energy/entropy content
\cite{BMS}.

The subtlety of the particle/field problem (not to be confused with the
particle/wave dualism of QM which was already solved by the transformation
theory of the 20s) was confirmed in the discovery of perturbative
renormalization and the time-dependent scattering theory\footnote{The elegant
formulation leading to the well-known useful expressions in terms of
correlation functions are due to Lehmann, Symanzik and Zimmermann (LSZ
formalism) whereas the proof of the asymptotic convergence towards free fields
was supplied by Haag and Ruelle \cite{Haag}.}. The main conceptual message was
that in interacting QFT the notion of particles at finite spacetime had no
intrinsic covariant (reference system-independent) meaning. Only the
asymptotic particle states are intrinsic and unique, whereas the fields (basic
or composites within the chosen description) form an infinite set of frame
independent objects whose physical nature is however somewhat fleeting since
observationally they carry a large amount of redundancy (infinitely many
different fields in the same local equivalence class lead to the same
asymptotic particle and scattering amplitudes. The situation resembles the use
of coordinates in geometry; the redundancy inherent in the use of different
coordinate systems corresponds to the use of different field coordinatizations
generating the same system of local operator algebras which correspond to the
intrinsic (coordinate-free) way of doing geometry.

This view is reflected in the terminology of the 50s, when fields were
referred to as "interpolating" fields, thus highlighting that they should be
considered as mediators of events involving particles without acquiring direct
observational relevance by themselves. In fact the algebraic approach, which
started shortly after the LSZ scattering theory, had as its main aim the
establishment of a setting in which the infinite plurality of fields is
encoded into the infinite ways of coordinatizing a \textit{unique}
\textit{system of spacetime-localized algebras}. In this way the setting of a
spacetime-indexed net of operator algebras represents a compromise between an
extreme on mass-shell/S-matrix point of view and a formulation in terms of the
infinitely many ways of generating the same unique net of spacetime-indexed
algebras using different covariant "field coordinatizations". \ 

This particle-field problem has again become a controversially debated issue
in the setting of QFT in curved space time (CST) \cite{Wa} when the
Poincar\'{e} symmetry including the notion of the vacuum and particle states
is lost. There are many results of QFT which are consistent with the
Lagrangian quantization setting (with which QFT is often incorrectly
identified), but which cannot be derived by textbook Lagrangian methods and
rather require operator algebraic methods. In this case it may be helpful for
the reader to replace the standard terminology QFT by local quantum physics
(LQP). The main difference is methodological and consists in the use of
field-coordinatization-independent algebraic methods wherever this is possible.

There exists an important area of QFT for which up to this day the use of
pointlike covariant field coordinates cannot be avoided namely
\textit{renormalized perturbation theory}. But even there the causal
perturbation theory a la Epstein-Glaser \cite{E-G} in terms of an iterated
lowest order input (in the form of an invariant polynomial pointlike coupling
between free fields) contains some of the LQP spirit. The coupling of free
fields to invariant interaction polynomals has hardly any direct relation to
Lagrangian quantization\footnote{The covariantization of Wigner's unique
representation theoretical classification leads to infinitely many covariant
(spinorial) fields (appendix), but most of them do not result from an
Euler-Lagrange principle. The latter may be necessary in starting from
euclidean functional integral representations, but they are not required in
the E-G setting.}. The method is based on the iterative application of the
causality and spectral principles of QFT; it does not follow the quantum
mechanical logic of defining formal operator as e.g. Hamiltonians via momentum
space cutoffs as unbounded non-covariant operators whose cutoff dependence
must then be removed in order to be formally consistent with the principles.
But even when the E-G formalism would reach its limits in the problem of
separating the infrared dibergencies from short distance problems of
perturbative renormalization of nonabelian gauge couplings, there is still the
possibility of a saving grace by separating the issue of states from operators
and operator algebras and in this way arrive at an infrared finite local
algebraic structure and leave the infrared problems to the construction of
states \cite{Fre}. Such a operator-state dichotomy is impossible in the
Lagrangian or functional integral formulation.

There was however one seemingly mysterious property in the particle-field
relation which, even using the advanced conceptional tool box of LQP, did not
reveal its mystery. This is the \textit{crossing property} (often called
misleadingly "crossing symmetry"). Only recently this property has lifted some
of its secrets (see last two sections). Since this property and other ideas
which resulted from it constitute the central subject of the present essay, a
clear definition is paramount. Fortunately this is not difficult since the
problem is not in its presentation, but rather its connection with the
principles of QFT.

Its formal aspects in Feynman's perturbative setting was obtained by combining
two observations: the invariance of certain families of subgraphs in the same
perturbative order under the conjugate interchange of incoming with outgoing
lines (the graphical crossing), and the less trivial mass shell
\textit{projection} of the connecting analytical path resulting in an analytic
relation \textit{onto the complex mass shell} between amplitudes describing
two different scattering processes. It is this second step of demonstrating
the existence of an analytic path on the complex mass shell linking the
backward mass shell defined by analytic continuation in formfactors with the
interchange $in\longleftrightarrow out$ and $particle\longleftrightarrow
antiparticle$ which (even in the setting of renormalized perturbation theory)
remains somewhat nontrivial.

According to the LSZ scattering theory collision amplitudes can be obtained
from formfactors, hence it is natural to formulate the crossing identity first
in this context. A \textit{formfactor} is a matrix elements of a field between
"bra" states, consisting of say n-k outgoing particles, and k incoming
particles in a "ket" state. Taking the simplest case of a scalar field $A(x)$
between spinless states of one species it reads%
\begin{align}
&  ^{out}\left\langle p_{k+1},...p_{n}\left\vert A(0)\right\vert
p_{1}...,p_{k-1},p_{k}\right\rangle ^{in}\label{cross}\\
&  =~^{out}\left\langle -p_{k}^{c},p_{k+1},...p_{n}\left\vert A(0)\right\vert
p_{1}...,p_{k-1}\right\rangle _{c.o}^{in}\nonumber
\end{align}
in words: the incoming 4-momentum on the mass ahell $p_{k}$ is "crossed" into
the outgoing $-p_{k}^{c},$ where the $c$ over the momentum indicates that the
particle has been crossed into its antiparticle and the -sign refers to the
fact that the formfactor is not between physical states but rather involves
the analytic continuation of one. The subscript $c.o$ (contractions omitted)
indicated that contraction terms of $p_{k}~$and the other $p^{\prime}s$ (inner
products) which are absent in the uncrossed configuration must be excluded
after the crossing. Since their structure is different from that of the
uncontracted leading terms, they can be easily separated from the main term.
This notational complication can be avoided if one formulates the crossing
relation directly in terms of free incoming/outgoing fields instead of
particles (section 5).

The relation (\ref{cross}) would be physically void if it would not come with
an assertion of analyticity which connects the unphysical backward mass shell
momentum with its physical counterpart. The (still unphysical) crossing
identity (\ref{cross}) together with the analyticity which connects backward
to forward momenta constitute the crossing property; there is no direct
identity between physical formfactors, only the affirmation that they are
related by analytic continuation. The proof is provided by modular
localization which will be the central issue in section 5.

The iterative application of the crossing relations permits to compute general
formfactors from the vacuum polarization components of $A(x)$%
\begin{equation}
\left\langle 0\left\vert A(0)\right\vert p_{1},p_{2},...,p_{n}\right\rangle
^{in}=~^{out}\left\langle -p_{k+1}^{c},...-p_{n}^{c}\left\vert A(0)\right\vert
p_{1}...,p_{k-1},p_{k}\right\rangle _{c.o}^{in}%
\end{equation}
where the charge conservation forces particles to be crossed into
antiparticles. Only the vacuum polarization matrixelement does not need the
subscript $c.o~$since contraction terms occur solely between bra and ket
momenta. The identity only holds for unphysical momenta. By analytic
continuation one can get to any formfactor with the same total number of
particles starting from the vacuum polarization component i.e. a local "bang"
on the vacuum $A\Omega$ determins all formfactors.

The S-matrix elements result from the formfactors by choosing for $A(x)$ the
unit operator\footnote{This could be achieved by cluster factorization of
$A(x),$ assuming that $A$ has a nontrivial vacuum component.}; since the
latter cannot absorb energy-momentum, the incoming momenta are bound to the
outgoing by the energy-momentum conserving delta function which leads to some
peculiarities. The analyticity in the momentum space representation can only
be valid for the function which remains after extracting the delta function.
Hence by crossing one particle it is not possible to return to a physical
scattering process. One needs to cross simultaneously an incoming and outgoing
pair in order to preserve the energy-momentum delta function for physical
momenta. This is particularly obvious if the crossing starts from a
two-particle state so that crossing only one particle will not lead to an
analytic continuation of a physical process. The formfactor of the identity
operator with the vacuum or with the one particle state on one side vanishes.
In order to come to a relation whose analytic continuation has a nontrivial
relation to elastic 2-particle scattering one must simultaneously cross a
particle from the opposite side i.e. cross a pair in exactly the way in which
crossing was first observed for two particle scattering in the setting of
Feynman graphs. We will return to this case in section 5.

Crossing looks as being closely related to TCP. Although both properties can
be derived from modular localization, the derivation of crossing turns out to
be much more subtle than that of the TCP theorem.\textbf{ }

The main conceptual role of crossing is that it relates the various n-particle
matrixelements of a local operator which belong to different distributions of
n particle momenta into incoming ket and outgoing bra states of an analytic
\textit{master function}. This is of course much more than the tautological
statement that these matrix elements can be computed once a concrete model has
been selected; it really means that once one process has been computed, the
others are uniquely determined in a model-independent way without doing
another QFT computation.

Since this essay also addresses readers with interests in philosophical
aspects, the occasional use of metaphoric arguments as a rapid vehicle to
convey a mathematically difficult property which places LQP into sharp
contrast with QM (even in its relativistic form \cite{Co-Po}\cite{interface}%
\footnote{The "direct particle interaction" (DPI) \cite{Co-Po} is a
relativistic particle setting which fulfills all properties of relativistic
particles which one can formulate in terms of particles only including
macro-causality (cluster factorization). Crossing can however not be
implemented in such a setting.}) should not cause problems. In any case this
will be limited to solved problems whose mathematical presentation can be
found in the existing literature.

The crossing properties of formfactors point at the most important consequence
of causal localization in the presence of interactions: the ability to couple
all multi-particle channels with the same superselected quantum numbers with
each other, in particular the non-orthogonality of corresponding localized
states. This is a double edged knife, it makes QFT much more foundational than
QM but it also renders many operator techniques one learns in QM unusable. In
the case of formfactors the analytic properties of crossing prevent that there
are special matrix elements which vanish, leading to the absence of certain
processes. Crossing is a special illustration of a general property of LQP
which often is expressed in an intuitive way as a kind of benign form of
"Murphy's law": \textit{particle states which (by charge superselection rules)
are allowed to communicate (via formfactors), actually do communicate i.e.
their coupling cannot be prevented, it rather constitutes a structural
property of any QFT. }It is this property which is behind the
interaction-induced (infinite) vacuum polarization clouds resulting from
"banging" with a local operator $A$ on the vacuum; but it is certainly less
metaphoric than the standard textbook presentation of the vacuum as a
"broiling soup of virtual particles" which is allowed to violate the
energy-momentum conservation for short times thanks to the uncertainty relation.

A special but important case of Murphy's law governing the coupling of
channels is the principle of \textit{nuclear democracy}. It states that QFT
cannot distinguish between elementary and bound particles, the only hierarchy
consistent with nuclear democracy is the already mentioned one between basic
and fused \textit{charges}. This means in particular that it is consistent to
view any particle as the result of a fusion of a cluster of other particles
whose fused joint charge is contained in the reduction of the fused charge
spectrum of the cluster under consideration. Nuclear democracy is certainly a
principle which contradicts the boundstate hierarchy of QM in a very radical
way; if even the charge-carrying "elementary" particle can be interpreted as
resulting from the collective fusion of its own charge with that of a local
cluster of suitably chosen other local charges, then the strict hierarchy
between elementary (fundamental) and composite certainly breaks down. Hence
regarding the formation of "bound particles" i.e. eigenstates of the mass
operator with a fused charge, the situation is \textit{radically different
from that in QM} because there is nothing which will prevent this particle
from coupling in a formfactor to all other states which the superselection
rules permit. The crossing property transfers the validity of \textit{Murphy's
Law} and the resulting principle of nuclear democracy for formfactors to the
phenomenon of vacuum polarization where there exist theorems showing that no
vacuum polarization component can vanish in a theory with nontrivial
interaction\footnote{The strongest result is a forthcoming theorem by Jens
Mund (private communication) which generalizes the old Jost-Schroer theorem
(see \cite{STW}).}. \ 

Let us now sketch the ideas which were used in the original proof of crossing
\cite{BEG}. The elastic 2-particle amplitude is a function of the 3 Mandelstam
variables s,t,u which are not independent but obey the relation $s+t+u=~m_{1}%
^{2}+m_{2}^{2}+m_{3}^{2}+m_{4}^{2}.$ There are 3 physical processes (and their
TCP conjugates) which can be reached if one knows the amplitude as a function
of the full range of the Mandelstam variables s, t. Bros, Epstein and Glaser
started from the LSZ representation in terms of Fourier transforms of time
ordered functions and used known analytic properties of the latter in order to
show that the physical region in terms of Mandelstam variables s%
$>$%
0, t%
$<$%
0 can be connected to the two other possible physical regions by an analytic
path. The proof is somewhat involved because it is not the primitive
analyticity domain of the starting correlation function but rather its
holomorphy envelop which leads to the desired result. Although the analytic
prerequisites for the continuation between forward and backward mass shell
through the complex mass shell are proven, the crossing identity does not
explicitely appear in that work.

These papers are an illustration of the profound mathematical knowledge which
physicists acquired in the pursuit of structural problems in QFT of the 60s.
The proof of crossing and later generalizations only addressed special cases
of scattering. However the interesting connection with thermal KMS properties
and a more general proof only came into the open in connection with formfactor
crossing as will be explained in the sequel.

Only decades later it became clear that localization in QFT (restriction of
the vacuum state to the local subalgebra) converts the vacuum state to a
thermal KMS state\textbf{ }%
\begin{equation}
\Omega_{vac}\upharpoonright\mathcal{A(O})\text{ }is\text{ }\Omega_{KMS}%
\end{equation}
where the Hamiltonian is canonically determined in terms of $(\mathcal{A(O}%
),\Omega).$ The mathematical theory behind this is \textit{modular theory}.
This theory exists in two interconnected versions, the operator algebraic
Tomita-Takesaki theory (of which important physical aspects were discovered
independently by Haag, Hugenholtz and Winnink \cite{Haag}) and the modular
localization of relativistic wave functions and states \cite{BGL}\cite{Sch}.
These ideas led to a closer connection of the thermal aspects of event
horizons in QFT in CST with thermal aspects caused by restricting the vacuum
state of global QFTs to localized algebras. A much discussed case is the
restriction of the vacuum to the wedge-localized algebra $\mathcal{A}(W)$
which leads to the Unruh effect and an interesting formula for the entropy
near the horizon $H(W)$ (the entropy of a light-sheet \cite{BMS}). The
$W$-Rindler word with its lightfront horizons is created in form of a
Gedankenexperiment involving a family of uniformely accelerated observers.
Black holes with their event horizons would lead to more real astrophysical
illustrations of thermal effects resulting from modular localization.

Two additional facts finally led to the somewhat surprising result that also
the crossing relation belongs to those phenomena which are related to thermal
aspects of localization. The first perception was the observation that at
least formally the KMS relation written for formfactors of free fields. For
free fields and their composites restricted to a wedge region (with the test
functions alway having support in W) one has\footnote{The reader should pay
attention to the changes of notation between expectation values and
matrixelements of operators between states.}%
\begin{align}
\left\langle A(f_{l+1})..A(f_{n})C(h)A(f_{l})..A(f_{1})\right\rangle  &
=\left\langle A(f_{1})\Delta A(f_{l+1})..A(f_{n})C(h)A(f_{l})..A(f_{2}%
)\right\rangle \label{id}\\
\left\langle p_{n}..p_{l+1}|C(h)|p_{l}..p_{1}\right\rangle  &  =\left\langle
p_{n}..p_{l+1},-p_{1}|C(h)|p_{l}..p_{2}\right\rangle _{c.o}\nonumber
\end{align}
Here $C(h)$ is a h-smeared composite of a free field. For the validity the KMS
relation with respect to the modular Hamiltonian $\Delta=e^{-2\pi K}$ with $K$
the Lorentz boost the smearing functions must be localized in $W$. Since the
mass-shell restriction of wedge-localized smearing functions form a dense set
of wave functions, the momentum space relation in the second line is a
consequence. The negative sign of the first momentum is a result of the
analytic continuation implied by the imaginary $2\pi$ Lorentz rotation
together with the Hermitian adjoint from passing from ket to bra states
(\ref{bra}); for obvious reasons the backward mass shell momenta are referring
to particles with the opposite charge i.e. "anti" with respect to the original
one before the cyclic permutation. To obtain the particle states from field
states one must Wick-order the $A$-field states on the left hand side and
remember that the cyclic permuted $A(f_{1})$ has no contractions with the
fields on the right from where it was coming. In the transcription of this
relation to particles, the absence of left backward momentum states contracted
with right forwards ones is indicated by $c.o.$ (contractions omitted).

Hence the crossing relation in the interaction-free case is noting else than
the thermal KMS relation of wedge localization (featuring in the Unruh effect)
rewritten as a relation between particle matrix-elements (formfactor). In
section 5 it will be shown that the interacting case is the particle
transcription of a \textit{new modular theory-based field relation which
extends the KMS relation}.

The correlation functions have analyticity properties in the Lorentz boost
parameter, they are analytic in the multi-strip \cite{Araki}%
\begin{align}
0  &  \leq\tau_{1}\leq\tau_{2}\leq..\leq\tau_{l}\leq\tau\leq\tau_{l+1}%
..\leq\tau_{n}\leq1\label{Ara}\\
A(f_{i})  &  \rightarrow e^{2\pi\tau_{i}K}A(f_{i})e^{-2\pi\tau_{i}%
K},~C(h)\rightarrow e^{2\pi\tau K}A(f_{i})e^{-2\pi\tau K}\nonumber
\end{align}
The support properties in x space of wave functions are equivalent to
analyticity properties. In particular they imply that certain complex Lorentz
transformations which act on the Fourier transformed operators can be absorbed
in the analytic continuation of test functions and vice versa.\textbf{ }

Looking only at the contribution of (\ref{id}) without contractions among the
free fields and using the density of Fourier transformed wedge-supported
smearing functions on-massshell, one obtains the crossing relation for the
free formfactor
\begin{equation}
\left\langle p_{n}..p_{l+1}\left\vert C(0)\right\vert p_{l}..p_{1}%
\right\rangle =a.c._{q^{c}\rightarrow-p_{1}^{c}}\left\langle q,p_{n}%
..p_{l+1}\left\vert C(0)\right\vert p_{l}..p_{1}\right\rangle _{c.o}
\label{test}%
\end{equation}

where the subscript $c.o$ has the same meaning as before\footnote{Instead of
omitting certain contration terms one might as well use the unmodified
formfactor and subtract terms of the form $c.t.=\sum_{r=l+1}\delta(p_{1}%
^{c}-p_{r})\cdot lower~formfactors$}. This is an identity between a particle
matrixelement of C and an a crossed formfactor at an analytically continued
momentum; (the notation $-p^{c}$ instead of simply $-p$ indicates that the
momentum on the backward shell is that of an antiparticle relative to what it
was before the crossing on the ket side. The only somewhat tricky part of
rewriting the KMS relation (\ref{id}) into the crossing form (\ref{test}) is
taking the operator $A(f_{1})\Delta$ as its conjugate to the bra vacuum and
using modular theory to bring the resulting bra state into the desired form
(for the notation see appendix)
\begin{align}
&  \Delta A(f_{1})^{\ast}\Omega=\Delta SA(f_{1})\Omega=\Delta^{\frac{1}{2}%
}JA(f_{1})\Omega=A^{c}(\check{f}_{1})\Omega=\int\frac{d^{3}p}{2p_{0}%
}\left\vert p^{c}\right\rangle \bar{f}(-p)\label{bra}\\
&  S=J\Delta^{\frac{1}{2}},~SA\Omega=A^{\ast}\Omega,~A\in\mathcal{A}%
(W),\mathbf{~~~}\check{f}(p)=\bar{f}(-p)\nonumber
\end{align}
More details can be looked up in section V.4 of \cite{Haag}%
\footnote{Especially recommended to philosophically motivated readers who
prefer conceptual clarity over mathematical rigor.}. The application of the
unbounded modular operators $\Delta^{\frac{1}{2}}=e^{-\pi K},K=W$-associated
Lorentz boost generator requires precisely that analytic continuability which
is guarantied by the wedge localization. With respect to analyticity there is
no difference between the KMS setting and its two-algebra generalization
needed for the derivation of crossing in section 5.

Instead of invoking modular theory, the free field relation (\ref{test}) can
also be checked by explicit computation, but this privilege does not exist in
the presence of interactions.

There is in fact a serious obstacle against applying this argument to
interacting formfactors in order to establish the identity (\ref{cross}). The
reason is obvious since there are 3 different algebras involved $A_{in}%
(W),A_{out}(W),A(W)$ and the modular operators of interacting operator
algebras are different from those generated by their asymptotic free fields.
But there is a fortunate circumstance which comes to one's rescue: at least
the domains of the unbounded Tomita S operators $S_{in},S_{out},S$ are
identical i.e. the $\Delta^{\prime}s$ coalesce and hence the dense subspace of
localized states are the same. The consequences of the identity of the domains
are the subtle ingredients in the proof of crossing. We will return to this
problem in section 5 and 6 and show that this suffices in order to derive
crossing in the formfactor- as well as in the scattering- form.

It is interesting to compare the old derivation \cite{BEG}\cite{E-G-M} which
uses holomorphy properties of correlation functions in several variables,
including the sophisticated tool of computing holomorphy envelopes (cutting
"noses"), with the present one. The wedge localization approach is quite
different, even though both rely on analyticity properties coming from
locality. Its analytic underpinning is that of Araki's KMS analyticity for
correlation functions and states \cite{Araki}.

The modular approach is more economical in the sense that only the analyticity
which is really necessary for the derivation of crossing is used, and analytic
completion techniques, whose physical interpretation is not known, are
avoided. In this way the important role of crossing in the construction of
factorizing models becomes clearer \cite{Sch}\cite{Lech1}. Finally crossing
becomes part of a structural problem of wedge algebras whose thermal
manifestations are important in the Unruh effect associated with a (Rindler)
wedge and its causal horizon as well as in thermal aspects related to event
horizons, including vacuum polarization-induced entropy near null-horizons
\cite{BMT}. This connection between properties from the center of particle
theory, with properties which at least historically come from black hole
physics, is the real surprise here.

At the time of the Bros-Epstein-Glaser work on crossing some quantum field
theorists pinned high hopes on the use of new analytic methods for functions
of several complex variables for a nonperturbative understanding of QFT.
K\"{a}ll\'{e}n and Wightman \cite{Ka-Wi} tried for many years to construct a
representation of the 3-point function which fulfilled all linear requirements
of QFT. They never reached their goal, and this kind of technique subsequently
fell out of favor.\textbf{ }Whether is returns one day together with different
problems, who knows? \textbf{ }

\section{The dual resonance model, superseded phenomenology or progenitor of a
new fundamental theory?}

The history of the crossing property starting in the early 60s is the key for
understanding the direction into which a good part of particle physics
research developed afterwards. It began by more or less accidentally stumbling
across a property whose importance in particular for an S-matrix-based
approach to particle physics was apparent, but whose foundational aspects
remained hidden. The necessary conceptual and mathematical tools for its
understanding only appeared at the end of the century in the form of modular
localization (appendix and sections 5,6).

Direct numerical attempts to find approximate solutions of the extreme
nonlinear properties of the the S-matrix bootstrap "axioms" ended in failure
but unfortunately only strengthened the misleading belief of the existence of
a unique non-Lagrangian theory of strong interactions. This was neither the
first nor the last time that an ultra reductionist "theories of everything"
(TOE) entered the particle theory discourse.

As mentioned before, after the completion of the dispersion theory project the
underlying philosophy of research began to change. The new strategy was most
clearly formulated by Mandelstam. In analogy to the rigorously established
Jost-Lehmann-Dyson spectral constructions for matrixelements of field
commutators \cite{Tod} (generalizations of the simpler K\"{a}llen-Lehmann
representation for the two-point function) which became a seminal tool in the
derivation of the dispersion relations, Mandelstam proposed an spectral
representation for the two-particle scattering amplitude \cite{Vech} in the
hope that the crossing property may be simpler accessed in terms of spectral
functions. This representation was never proven and the hope about its use did
not materialize, but taken together with ideas about the use of Regge pole
trajectories in strong interaction phenomenology it led Veneziano to the
mathematical construction of the dual resonance model for elastic two-particle
scattering\footnote{The added "resonance" expressed the wish to unitarize the
model so that it could pass as an S-matrix Ansatz.} \cite{Vech} which was
later generalized to an arbitrary number of particles.

In terms of Feynman graph terminology it represented the tree approximation
for a process of two incoming particles which couple via trilinear interaction
vertices to an infinite tower of intermediate particles with ever increasing
masses and spins. The decrease of the coupling strengths is carefully tuned in
such a way that the sum of all these contributions from the infinite mass/spin
tower of the interaction mediating particle poles not only converge in the
s-channel (using the canonical terminology introduced by Mandelstam), but
represents a function which allows a t-channel interpretation in terms of
another sum of infinitely many exchanges via particles from the same mass/spin
tower. To find such function in a pedestrian manner, without an operational
backup, just by using known properties of gamma and beta functions, is an
astonishing achievement which even nowadays commands respect \cite{Vech}.

In hindsight it is somewhat surprising that it was not realized that the dual
model and its Nambu-Goto Lagrangian string theory analog which share the same
particle/spin tower were the first nontrivial realizations of an object which
less than one decade earlier was searched for under the label \textit{infinite
component fields}. The motivation came from a completely different corner,
namely from the analogy to the $O(4,2)$ "dynamic symmetry" of the hydrogen
atom. Infinite component fields in the sense of Fronsdal, Barut, Kleinert and
other authors \cite{Tod} were not just infinitely many fields of varying mass
and spin put together as a direct sum, but there was a "dynamic" content
consisting in the existence of operators which "vertically" communicate
between the different tower levels and set the mass/spin spectrum. This
dynamic aspect was expected to arise from noncompact group representations
which extend those of the Lorentz group, but this hope did not materialize and
the cited authors remained empty-handed. This dynamic requirement makes the
construction of an infinite component field a difficult problem. In fact up to
date the 10-dimensional superstring field has remained the only dynamical
infinite component pointlike solution in which the representation of the
Poincar\'{e} group is a positive energy unitary ray representation.

String theory owed its social success as an infinite component field theory
only 6 years after the ill-fated infinite component program to the replacement
of higher noncompact groups by the infinite degrees of freedom inherent in
multicomponent chiral conformal currents or equivalently in a canonical
quantization of the bilinearized (square roots removed) Nambu-Goto Lagrangian.

It is one of the missed chances in history that even though the followers of
the infinite component field program and those of the later dual model
community (which afterwards became incorporated into the string community) had
both strong phenomenological roots, they never noticed the proximity of their
ideas. It certainly would have been very interesting to be informed that the
duality requirement imposed on the vertices of a pole approximation for a
scattering amplitude and the mass/spin tower of the Nambu-Goto description of
string theory can be encoded into an infinite component field containing
operators which intertwine between the levels of the infinite tower. The word
"string" would never have appeared and there would have been no danger in
misreading string theory as having something to do with string-like objects.
Since one cannot change 50 year old customs, the word string theory will
always refer to this infinite component pointlike field, whenever we talk
about real string (e.g. in gauge theories) we will use the terminology "string-localized".

The duality idea arose from consistency arguments between the low energy
resonance contributions and the expected high energy Regge behavior.
Veneziano's first implementation led to several generalizations \cite{Vech}.
The formulation which is most suitable for an in-depth critical analysis is
the operational setting of Fubini et al. \cite{Fu-Ve} which uses
multi-component conformal currents and their potential.

It may be helpful for the reader to recall at this point some results about
conformal currents \cite{Vech}. The simplest situation is that of a
one-component current which, similar to a free field, is determined by its
commutation relation%
\begin{align}
&  \left[  j(x),j(y)\right]  =-\delta^{\prime}(x-y)\\
Q  &  =\int j(x)dx,~~\psi(x)=~"e^{i\alpha\Phi(x)}",~~\Phi(x)=\int_{-\infty
}^{x}j(x)dx \label{int}%
\end{align}
Despite its simplicity it leads to a very rich representation theory. There
are continuously many representations (labeled by $\alpha)~$as a consequence
of the continuous spectrum of the charge Q \cite{BMT}, so in the jargon of
chiral QFT these models are extremely "non-rational". Formally such charged
fields are written as exponentials of "potentials" i.e. half space integrals
over the current. The quotation marks are meant to indicate that such formulas
are conceptually not quite correct since the charge $\alpha$ carrying field
$\psi$ does not live in the vacuum sector as the naive reading of this formula
would indicate\footnote{If one uses such formulas outside of the theory of
superselected charges one must add the charge conservation by hand; only then
does one obtain a Wightman theory in a Hilbert space.}. This observation is
inexorably linked with the infrared divergence of the integral representation,
which is the way in which the exponential announces that it is not a quantum
field like the others in the Hilbert space generated by the currents.
Unfortunately the extended algebra which incorporates \textit{all}
charge-carrying fields lives in an inseparable Hilbert space.

In order to use currents as a two-dimensional theoretical laboratory following
the intrinsic logic of QFT, Buchholz, Mack and Todorov \cite{BMT} introduced
the concept of \textit{maximal local extension} of the algebra of currents.
The extension is done by adding certain fields of the form $\psi_{\alpha}(x),$
whose dimension $d_{\alpha}\sim\alpha^{2}$ is integer (and hence which for
different localization points commute among each other) to the algebra of
currents and view the resulting larger bosonic algebra as the \textit{extended
observable algebra}. This reduces the number of charge sectors in a drastic
way, their number is now not only countable but even finite ("rational chiral
theories"). It turns out that the denumerable set of maximal extension can be
explicitly constructed. These do not commute among themselves or with each
other but rather obey (abelian) braid group commutation relation.

The multi-component generalization of the representation theory of a current
turned out to lead to a theory of remarkable richness \cite{Stas}\cite{Longo}.
In this case the maximal extensions are classified by \textit{even lattices}
$L$ in $\mathbb{R}^{n},~L:$ $\left(  \alpha,\beta\right)  =2\mathbb{Z}$.$~$The
sectors are then classified by equivalence classes of the dual lattice
$L^{\ast}/L$ of which there exist finitely many. The cases with $L=L^{\ast}$
are particularly interesting$.$ These constitute a finite number of models
which only exist in their vacuum representation. They are related to finite
exceptional groups, among them the famous "moonshine model". This is just to
mention that the following operator dual model construction happens in a
fascinating neighborhood.

Besides this use of multicomponent current models following the intrinsic
logic of LQP, these currents have also been used in an operational approach to
the dual model in the work of Fubini at al. \cite{Fu-Ve} which is somewhat
different from the conformal field theoretic logic. Their interest was in the
direct use of the potentials $\Phi_{i}$ of the multi-component currents as
some quantum mechanical objects%
\begin{align}
\Phi_{i}(x)  &  =\int_{-\infty}^{x}j_{i}(x)dx\rightarrow X_{i}(z),~i=1....d\\
Q_{i}  &  \rightarrow P_{i},~\alpha_{i}\rightarrow p_{i},\text{ }%
V(z,p)=e^{iP\cdot X(z)}\nonumber
\end{align}
this symbolic formulas are in need of some detailed explanation. The first
line indicates a passage from the noncompact to the compact picture
$x\rightarrow z,$ and the notation $X_{i}(z)$ anticipates that the potentials
are now going to be interpreted as quantum coordinates which classically would
trace out a curve in a d-dimensional spacetime. The second line expresses the
fact that one really wants to take this reinterpretation into a different
direction by adding the identification of the d-component charge operator with
the momentum operator and writing the charge-carrying exponential of an
would-be n-component "spacetime" as a vertex (or in more recent
generalizations \textit{a chiral sigma field}) operator $V$ which carries a
noncompact spacetime symmetry (which from the chiral conformal viewpoint of
the source theory would be called an inner symmetry).

This is the famous source-target relation which later led to the notion of
world sheets. But is this strange interpretation of multicomponent charge
values as momenta with the operator dimensions of the charge-carrying
operators passing to particle masses and the current potential $\Phi_{i}$
becoming a kind of position operator in an multicomponent internal symmetry
space consistent? Is such a magic of defining a "target" spacetime in a
source-target relation supported by any physical concept? This picture would
suggest that the potentials of the conformal current theory define an
embedding of a line/circle into spacetime which in turn is the origin of the
worldsheets (in analogy to Feynman's worldlines) in the string theoretic
extension of the dual model. Admittedly the identification of an internal
symmetry space with noncompact physical spacetime is one of the strangest
ideas which entered particle physics \footnote{In higher dimensions it has
been shown from first principles that all inner symmetries are described by
compact groups \cite{Do-Ro}.}, but is it consistent? As oftern, the idea is
abscured by the very inappropriated terminology "field space" which somehow
insinuates that the space in which classical fields take their values
continues to make sense in QFT. We will use instead "internal symmetry space"
which is probably what is meant. We will show in the following that the
worldsheet picture is incorrect and that instead the localization is pointlike
as in standard QFT, which leads to worldlines and is consistent with what was
said before about ST really dealing with infinite component pointlike fields.

Fact is that one cannot embed any lower dimensional QFT into a higher
dimensional one. The frame-independent modular localization of QFT as opposed
to frame-dependent Born-Newton-Wigner localization of (relativistic) QM, is a
totally holistic concept which does not allow such embeddings which in QM
would be a triviality. The degrees of freedom of the theory to be embedded
never go into a stringlike extension of localization, they rather form their
own space which manifests itself as an internal infinite component space. A
quantum mechanical construction does not know where it is going to live, the
acting designer has to tell where he wants it.

One can only restrict a higher dimensional QFT to lower dimensional part of
spacetime. But physically this is generally not a very useful procedure
because the lower dimensional theory obtained in this way will inevitably have
too many phase-space degrees of freedom for being a candidate for a physical
QFT in that lower dimension. The only exception to this rule is the
holographic projection onto a null-surface resulting from a causal or event
horizon. A AdS$_{5}$-CFT$_{4}$ correspondence or a restriction of a QFT to a
brane is mathematically possible but only with an unphysical abundance of
phase space degrees of freedom on the side with the lower dimensions or a
unphysical anemia in the opposite direction. The unphysical consequences have
been studied by people who have studied QFT beyond its Lagrangian confines
since in that case it is impoetant to know which property guaranties the
causal propagation (this is much more than the spacelike commutativity) and
the existence of temperature states for arbitrary temperatures.

In order to discuss problems of unitarity of Poincar\'{e} representations on
inner symmetry space of chiral theories it is inconvenient to use the dual
model setting. The reason is that even in case of the 10 dimensional
superstring for which Hilbert space- as well as energy- positivity can be
satisfied, the supersymmetric unitary representation is only obtained after
passing to a subspace and dividing out zero norm-states. This blurrs the
picture of the target spacetime resulting from an inner symmetry of a
multicomponent potential of a current, and the presentation in terms of the
bilinearized Nambu-Goto Lagrangian (see next section) is more convenient.

As indicated before one replaces the superselected charges by superposable
momenta and the potentials by operators $X_{i}(z)$ whose lowest Fourier mode
(which includes a logarithmic contribution) $X_{i}(z)_{0}=x_{i}^{op}%
+icp_{i}^{op}\ln z$ defines quantum mechanical $x^{op},~p^{op}$
operators\footnote{The appearance of the logaithmic term is a mark of the
formal infrared divergence of the potentials which by themselves (outside
their exponential form) are not conformal fields. The label op distinguishes
quantum mechanical operators from the numerical momentua (alias chage
values)}. In this way the inseparable Hilbert space which describes charged
representations for a continuum of charges is avoided and the continuous
direct sum becomes a quantum mechanical direct integral in the sense of
spectral decomposition theory. Although the presence of these quantum
mechanical degrees of freedom prevent the conformal covariance of the zero
dimensional $X_{i}(z)$ field$,$ there is no problem with the covariance of the
exponential vertex operators which carry an anomalous dimension proportional
the square of charges which in the new reading corresponds to the square of
momenta i.e. of masses%
\begin{equation}
d_{\psi}\sim\alpha\cdot\alpha,~d_{V}\sim p\cdot p=m^{2}%
\end{equation}

So in the Fubini et al. formalism \cite{Fu-Ve} Veneziano's rather involved
gamma function setting is replaced by a formalism using the conformal
invariant part (the part which depends only on the anharmonic ratios) of the
4-point function of the vertex operator. The higher point function dual model
amplitude results from the invariant part of the higher correlations; in this
way one arrives at a dual model representation for $n\rightarrow m$ particle scattering.

It is hard to criticize a proposal which is phenomenological in nature, apart
from expressing some unease about putting together raw phenomenology ideas
(which were later contradicted by new experiments) with subtle mathematical
concepts which already had a different very precise conceptual position. It is
probably the attractive mathematical aspect which explains why this proposal
did not disappear completely together with the Regge phenomenology when the
latter came to an end. Being a somewhat too ambitious setting for a mere
phenomenological description, the theory had its later comeback in the form of
string theory; but whereas its mathematical entitlement was natural, the same
cannot be said about its physical interpretation. It finally became acclaimed
as the millennium TOE which, different from the S-matrix bootstrap, allegedly
also includes gravity.

The conceptual distinction resulting from the of apparent uniqueness of
mathematically ambitious projects as the implementation of the highly
nonlinear duality structure has often mislead people\footnote{The nonlinear
S-matrix bootstrap and the Schwinger-Dyson illustrate this point.}. In the
beginning there was only Veneziano's version of the dual model which was
constructed by a clever use of properties of gamma functions. But now we know
that there are myriads of functions of the Mandelstam variables $s_{ij}$ which
are meromorphic with an infinite tower of particle poles in the position of
duality. They are constructed by starting from any conformal theory in any
spacetime dimension. As explained in detail in a beautiful paper of Mack
\cite{Mack}, one only has to write the connected part of a conformal n-point
function as a Mellin transform $M$%

\begin{equation}
G^{c}(x_{1},...x_{n})=\left(  \frac{1}{2\pi i}\right)  ^{n/2}\int..\int
d\delta M^{c}(\left\{  \delta_{ij}\right\}  )%
{\displaystyle\prod\limits_{ij}}
\Gamma(\delta_{ij})\left(  -\frac{1}{2}x_{ij}\right)  ^{-\delta_{ij}}%
\end{equation}
There are as many integration variables as there are independent conformal
invariant anharmonic ratios. The aim is to show that by identifying the
operator dimensions of the conformal fields with the masses of particles and
the Mellin variables $\delta_{ij}$ to the Mandelstam variables $s_{ij}$ one
obtains a meromorphic Mellin transform which has the correct poles as required
by the duality property. The reduced Mellin transform $M^{c}$ can be defined
in such a way that the spacetime dimensionality does not enter\footnote{The
properly reduced Mellin amplitudes are independent of spacetime dimensions;
this is similar (actually closely related) to the invariant part of conformal
correlation which only depends on dimension-independent conformally invariant
harmonic ratios.} i.e. one can obtain dual models in a fixed spacetime
dimension from conformal theories in \textit{any} dimension, not only from
chiral conformal theories.

The convergence of the infinite sums over poles as well as certain positivity
properties of the associated residues follow from the established validity of
\textit{global operator expansions in conformal theories} \cite{Mack}. At this
level, there is however no claim that the Mandelstam variables are related to
momenta on which a unitary representation of the Poincar\'{e} group acts. This
problem was not part of the dual model program since the only positivity
requirement in the Mandelstam setting of scattering amplitudes are conditions
on the correct sign of residua of poles. It however became a pressing problem
after the original phenomenological purpose of the formalism was abandoned and
the setting was allowed to become the driving force of a free-roaming TOE
under some mathematical, but practically no conceptual control. The rallying
point for this development was the observation that the only unitary positive
energy representation of a Poincar\'{e} group which can act on\ the index
space of a multi-component current and its potentials or on the oscillator
space of the Nambu-Goto model is the 10-dimensional superstring
representation. In this case the Mandelstam invariants result from a unitary
momentum space representation of the Poincar\'{e} group.

In the present context the Mellin formalism demystifies Veneziano's
observation to some extent in that it shows that the duality structure, far
from being a lucky discovery of a special way to implement (an approximated
form of) the crossing property, is in reality a kinematical aspect of a
certain transformation property of conformal correlation functions. Unlike the
Fourier transform of correlation functions it cannot be expressed in terms of
single operators but needs the entire correlation function for its
definition\footnote{One needs the conformally invariant part of the
correlation, a step which permits no operator formulation.}. The operator
version of the Veneziano dual model \cite{Fu-Ve}, which starts from a chiral
conformal current model, turns out to be a special case of Mack's conformal
Mellin transformation formalism. But whereas in the former the momenta enter
explicitly via the continuous charge spectrum, the appearance of momenta in
Mack's setting is less overt; they only enter in parametrizing a relation
which links the anomalous dimension of the conformal theory to the independent
variable in the Mellin transform \footnote{The interpretation of the
(appropriately defined) Mellin transform as a 4-dimensional dual model is
idependent of the spacetime dimensionality of the associated conformal model.
For the Fubini et al. \cite{Fu-Ve} model it is a multi-component abelian
chiral current.}.

As mentioned before the existence of a unitary positive energy representation
of the Poincar\'{e} group behind the Mandelstam variables is not part of the
Mellin transformation formalism. The verification of its existence in d=10
(the superstring theory) is certainly an unexpected curiosity since there was
no reason at the beginning to expect a chiral conformal theory to support a
noncompact inner symmetry as a Lorentz group representation start\footnote{The
use of inner symmetry indices of a QFT as an arena for representations of
spacetime symmetries is one of the strangest proposals ever made in particle
physics. Once accepted, it opened the flood gates for other metaphoric ideas
as e.g. the conversion of unwanted spacetime dimensions via "compactification"
into inner inner symmetries.}. But to take such a property of a
two-dimensional conformal theory as a hint of leading to a new understanding
about spacetime is far-fetched if not a step into mysticism.

As mentioned before, the infinite component field of superstring theory in
d=10 is the first and only nontrivial realization of a dynamic pointlike
irreducible infinite component theory in the before explained sense
\cite{Tod}. The protagonists of the infinite component field idea (if some of
them are still around) would perhaps notice with satisfaction that by allowing
quantum mechanical oscillators to connect the levels and to generate the
mass/spin spectrum one obtains the first illustration of what they had in
mind; perhaps they would have been less than happy about the high spacetime
dimension of this unique realization and its resulting metaphoric epiphenomenon.

A relation between masses and operator dimensions which is not related to
Mellin transformation occurs in a more intrinsic physical context of the
AdS-CFT correspondence. This correspondence will appear in a different context
in the concluding remarks.

Whatever one wants to make out of the operator setting of the dual model or
the Mellin formalism, there is certainly no intrinsic physical reason why one
should re-interpret charges as momenta and inner symmetry spaces of chiral
theories as spacetime arenas for physical events. And why should one follow
somebody who claims that the generating objects of ST are stringlike, in
blatant contradiction to the pointlike computational results which lead to
worldsheets on such an incorrect metaphoric path? The systematic construction
of dual model amplitudes via conformal QFTs has nothing to do with the
physical picture of an imagined approximation to the conjectured Mandelstam
representation, nothing can conceptually be farther apart than scattering
theory of particles and conformal QFT.

Why mystify the different 10 dimensional pointlike superstrings and their
presumed connection via M-theory as revealing deep secrets of physical
spacetime when there is the autonomous possibility of explaining these
properties as peculiarities of inner symmetries of chiral models which are
known not to have to follow the standard inner symmetry pattern in terms of
compact group representation of higher dimensional symmetries? Behind all this
is the general question: is particle physics only interesting after, following
the modern Zeitgeist, it has been sexed up or mystified?

\section{String theory, a TOE or a tower of Babel within particle theory?}

String theory addresses some of the questions which the dual model left open
or could not handle convincingly as: can one really obtain a unitary
representation of the Poincar\'{e} group on the internal symmetry space of a
chiral current theory and if yes, what is the covariant localization concept
in such a source-target relation and in particular does it really lead, as
claimed, to a notion of world sheets? Last not least one would like to know
whether the use of special exponentials of potentials (in the operator duality
approach) can be replaced by a more general setting in which, similar to the
Wigner approach to particles, a representation space is defined in terms of
generating wave functions with clear localization properties, which are then
used to pass to an (interaction-free) operator field formalism. For this
purpose it has turned out to be covenient to start from a slightly more
general point of view which prepares the desired unitary representation theory
more directly in terms of the current potentials $X^{\mu}(z).$

But before going into these technicalities some general remarks are in order.
There exist operator algebras and state spaces which have no pointlike but
rather semiinfinite string-like generators; Wigner's massless \textit{infinite
spin} representation family presents the only noninteracting illustration
\cite{MSY} and it shows that string localization is incompatible with a
Lagrangian description. In this case one may speak of world sheets being
traced out in spacetime. But the generating wave function of string theory and
their second quantized counterparts are pointlike generated. Originally the
string world sheets were not part of the dual model of old, they appeared in a
later stage when it was incorrectly claimed that the source-target relation
can be understood as an embedding of the one-dimensional chiral theory as a
one-dimensional submanifold into a 10-component target space representing
spacetime. To support such a picture string theorists invented a Lagrangian
description of relativistic particles \cite{Pol}. Compared with Wigner's clear
representation-theoretical classification, the functional integral
representation in terms of relativistic particle mechanics falls short of a
convincing attempt to support string theory; it is mathematically
ill-defined\footnote{It requires to pass through apparently unavoidable
infinite intermediate steps resulting from the necessity to extract infinite
factors coming from reparametrization invariance which have nothing to do with
intrinsic properties of particles.}, does not describe all irreducible
positive energy representations, and was never used by particle physicists who
characterize particles following Wigner. Such ad hoc inventions whose only
purpose is just to make one point in an analogy sometimes backfire instead of
lending support.

Before going into ST details, it is helpful to start with a theorem from
unitary representation theory which limits the localization of states (appendix).

\begin{theorem}
The causal localization (modular localization, see appendix) inherent in
unitary positive energy representations of the covering of the Poincar\'{e}
group is pointlike generated apart from Wigner's massless infinite spin
representation whose optimally localized generators are semiinfinite spacelike
strings \cite{MSY}.
\end{theorem}

Some comments are in order.

Unitary positive energy representations are canonically related to free fields
or (in case of reducible representations) to direct sums of free fields. The
bilinear Nambu-Goto Lagrangian is interaction free, hence the localization is
completely determined by the representation theoretical content. One only has
to show the absense of the Wigner infinite spin representation from the
positive energy unitary 10-dimensional superstring representation (which is
obvious) in order to secure that it is pointlike generated where pointlike
generated means that there is a collection of infinite component singular
functions\footnote{The different wave functions are distinguished by different
relative strength with which the different irreducible components contribute
to the mass/spin tower. String theory provides operators which change this
decomposition.} (wave function-valued distributions) $\psi(x)$ whose smearing
with test functions generate the one string space. In the Fock space extension
this corresponds to a collection of infinite component pointlike fields whose
one-string projection leads to the singular wave functions.

This theorem also covers the localization in string theory, since the
Lagrangian which underlies the quantum string is bilinear and hence the
(graded) commutator must be a c-number. This Lagrangian supplies the operator
formalism acting in the Hilbert space of the string wave functions. This
one-string representation space is an analog of the Wigner one particle space
apart from the fact that there is a severe restriction from the unitarity of
the action of the Poincar\'{e} group. This is because the central issue is the
quantization of a Lagrangian and the unitarity problem is an additional
restriction The situation resembles vaguely that of the vector potentials in
QED in that one has to form sub- and factor- spaces in order to get rid of the
negative and zero norm states. But whereas in QED this idea is independent of
the spacetime dimension and certainly does not effect the noninteracting
theory (where it only appears if one uses potentials instead of field
strengths), the origin in string theory is quite different. It can be traced
back to the unmotivated (i.e. not physically justifiable) demand that one
wants a \textit{unitary positive energy representation of the covering of the
Poincar\'{e} group on an internal symmetry space of quantum theory. }.

Nature could have answered this extravagant requirement by providing the same
negative response which has been known in higher dimensional QFT namely: any
inner symmetry is necessarily described by a compact group; noncompact groups
as spacetime symmetries would be in contradiction with the localization
principles of LQP \cite{Haag}. But surprisingly there are exceptions in chiral
QFT where besides "rational" models (which are in many ways similar to the
inner symmetry structure of higher dimensional models) and models with
countably many superselection sectors, there are also quite different
"irrational" internal symmetries. Models in which the observable algebras are
defined by multicomponent abelian currents belong to the latter. They have a
continuum of charged representations and there is indeed the possibility to
have (in an appropriate sense) a positive energy representation of the
covering of the Poincar\'{e} group on a 10 dimensional internal symmetry space
of a chiral current model. But from the context in which this somewhat
surprising observation arises it is clear that it has nothing to do with a new
mysterious insight into foundational problems of spacetime but rather with an
unexpected property of the particular chiral model (other surprising
properties of maximal extended current algebras were mentioned in the previous section).

Whatever one's position is towards spacetime symmetries appearing on the inner
symmetry space of chiral currents, there can be no doubt about the fact that
the one string space (or the uniquely associated string string field theory)
is pointlike generated. This is the unavoidable conclusion from the previously
stated theorem as well as from the below mentioned concrete calculations.

At this point it is important not to equate the localization of states with
that of operators beyond the setting of free fields. Whereas only the family
of massless infinite spin Wigner representations is semiinfinite string-like
generated \cite{BGL}\cite{MSY}, the absence of pointlike algebraic generators
in certain charged subalgebras is quite common. The best known case is that of
electrically charged fields in QED \cite{Jor}, it is impossible to localize a
charge-carrying operator in a compact spacetime region. Within massive
theories the possibility of such a situation was investigated by Buchholz and
Fredenhagen \cite{Bu-Fr}, but since in this case there would be no infrared
manifestation of string localization in Lagrangian perturbation theory, there
are presently no illustrative models of string localization in theories with
mass gaps \footnote{The presence of zero mass photons with an infrared-strong
coupling to charged particles results in a weakening in localization of the
latter. The optimal (sharpest) localization of the latter is semiinfinite
stringlike as described by the well-known Dirac-Jordan-Mandelstam line
integral representations. Charged fields interpolate "infraparticles" instead
of Wigner particles}. A B-F stringlike or an electrically charged field
applied to the vacuum decomposes into pointlike generated wave functions, but
this decomposition process has no counterpart in the local algebras.

By leaving the issue of localization in string theory to be settled as a
special consequence of a powerful structural theorem in local quantum physics
as above, one deprives oneself of some interesting insight into one of the
most fascinating episodes in 20th century particle physics namely a more
detailed understanding of where did the arguments leading up to string theory
fail. For this reason we will now follow this more interesting path.

The formal starting point is the bilinear Lagranian form in which the
Nambu-Goto Lagrangian \cite{Nambu}\cite{Goto} is used in string theory%
\begin{align}
L  &  =%
{\displaystyle\iint}
(\partial_{\tau}X_{\mu}\partial_{\tau}X^{\mu}-\partial_{\sigma}X_{\mu}%
\partial_{\sigma}X^{\mu})d\tau d\sigma\\
&  \left(  \partial_{\tau}^{2}-\partial_{\sigma}^{2}\right)  X^{\mu
}(z)=0\nonumber
\end{align}
In the simplest case the $\tau,\sigma$ dependent "zero scale dimension
position field" $X_{\mu}(\tau,\sigma)$ (the string analog of the Fubini...
potential) is considered to be defined on $R\times(0,\pi)$ with appropriate
(Neumann) boundary conditions. The equation of motion is a two-dimensional
wave equation which together with the boundary conditions leads to the Fourier representation%

\begin{equation}
X^{\mu}(\tau,\sigma)=x^{\mu}+p^{\mu}\tau+i\sum_{n\neq0}\alpha_{n}^{\mu
}e^{-in\tau}\frac{\cos n\sigma}{n}%
\end{equation}
The $\alpha_{n}^{\mu}$ are oscillator-type creation and annihilation operators
which by Lorentz covariance are forced to act in an indefinite metric space.
Denoting these chiral current potentials by $X^{\mu}$ may create the delusion
that one is describing a path in target space; with the conservative notation
$\Phi^{\mu}$ such an association is less automatic.

In the present form there is yet no free go for a unitary Poincar\'{e} on
target space, such a move must be more carefully prepared. Imposing subsidary
conditions%
\begin{equation}
\left(  \partial_{\sigma}X\pm\partial_{\tau}X\right)  ^{2}=0
\end{equation}
does the job, after they have been adjusted to the quantum setting (valid only
on states). The Klein Gordon equation on target space with a mass operator of
an integral spaced spectrum. These conditions express reparametrization
invariance and they would have been a consequence of the true Nambu-Goto
Lagrangian which is a nonlinear expression in the $\partial X.$ Clasically the
true N-G Lagrangian is equivalent to its bilinearized version plus
constraints. The reparametrization invariance trivializes part of the infinite
dimensional conformal covariance. All these aspects are subordinated to the
construction of a unitary Poincar\'{e} group representation on the
appropriately defined target space of a multicomponent current potential; they
do not have any intrinsic physical meaning.

Any unitary representation of the Poincar\'{e} group acts in a Hilbert space
can be obtained by a two-step process from a formal covariant representation
in a negative metric space of the form
\begin{equation}
H_{sub}\subset L^{2}(\mathbb{R}^{n},\rho(\kappa)d\kappa d\mu(p,\kappa))\otimes
H_{QM}%
\end{equation}
where the first factor is a spinless relativistic particle representation
space with a continuous mass distribution and $H_{QM}$ contains vector-valued
or spinor-valued quantum mechanical variables (as the $\alpha_{n}^{\mu}$)
which strictly speaking are prevented by Lorentz covariance to be genuine
"quantum" (acting in a Hilbert space).

In the simplest case of finite dimensional massive representations, the
representation space is the n-dimensional (nonunitary) vector representation
space of the Lorentz group $H_{QM}=V^{(n)}.$ To get to a unitary massive
$s=1~$representation of the Poincar\'{e} group one uses Wigner's idea of the
little group and obtains a \textit{unitary p-dependent Lorentz transformation
law} which results from the original non-unitary covariant law through an
\textit{intertwiner} (a 4-component function on the forward mass shell)
between the original n=4 vector representation with and its manifestly unitary
form which acts covariantly on a positive metric subspace $H_{phys}%
=H_{sub}\subset L^{2}(R^{n})\times V^{(n)}.$

In the N-G case at hand the selection of the mass specrum is done by imposing
the Klein-Gordon equation with the mass operator, its spectrum then leads to a
direct sum over equally spaced mass eigenstates including a "lowest" tachyonic
contribution%
\begin{equation}
\sum_{\kappa=-2,0,2,...}L^{2}(\mathbb{R}^{n},d\mu(p,\kappa))\otimes H_{Osc}%
\end{equation}
And one has to still implement the complete set of subsidery conditions. For
this purpose one uses the vector-valued oscillators belonging to the higher
Fourier components of the current potential whose Lorentz invariant inner
product is indefinite. There is no chance to find a subspace through
subsidiary conditions which is positive semidefinite with one exception. Only
for the multicurrent model with 26 components does one arrive at a
semidefinite metric \cite{Brower}. The last step is canonical, having arrived
at a semidefinite situation, the positive definite situation is gratis.
Details can be found in many articles \cite{Dim}. The obtained 26 dimensional
representation is not of positive energy as a result of the presence of a
tachyon. However admitting spinorial-valued chiral current components (which
would require a spinorial change of the N-G Lagrangian) one arrives at the 10
dimensional positive energy superstring representation.

The transition from unitary to covariant representations is done with the help
of so called $u,v$ intertwiners. This step is in complete analogy to what
Weinberg presented in a group theoretic setting in the first volume of his
well-known textbook \cite{Weinberg}. It also can be obtained by applying
modular localization to the Wigner representation theory \cite{MSY} (see
appendix). Although for each irreducible unitary representation there is only
one wave function space, there are infinitely many different looking covariant
wave functions and free fields (see appendix).

Since a unitary representation of (necessarily noncompact) spacetime symmetry
group on an internal symmetry space of a current algebra is a strange
requirement from a viewpoint of local quantum physics\footnote{The unresisted
acceptance of identifying inner symmetries of conformal symmetries with actual
spacetime and its opposite of mutating spacetime dimension into inner
symmetries by "rolling them up" (compactification) is an indicator for how
much the conceptual framework of QFT principles has been lost and replaced by
a collection of computational recipes.}, it would be very natural to have
received a negative answer to the target space issue. But inner symmetries in
low dimensional QFT are different from their standard realization and lo and
behold there is precisely one exception namely the positive energy
"superstring" representation in 10 spacetime dimension.

But does the existence of this exception indicate some mysterious new insight
into spacetime? Certainly not, it does however reveal some unexpected property
of the potentials $\Phi^{\mu}$ (and their charge-carrying exponentials) of
multicomponent chiral currents. Actually the solution is not completely unique
since there is a finite number of 10 dimensional superstrings and there exists
even a conjecture (M-theory) about their possible relations. It would be
interesting to present these observations (in analogy to Mack's Mellin
formalism) solely in terms of multicomponent currents and their potentials,
leaving spacetime metaphors aside.

The correct reading of the string as a dynamic infinite component
field\footnote{One has less problems with looking at the source \ --%
$>$
target embedding as a purely formal device.} shares the inner
symmetry$~\rightarrow~$spacetime symmetry reinterpretation with that of string
theory. But there is less temptation to elevate the construction of a
(possibly unique) dynamic infinite component field to a new foundational
insight into spacetime or to interpret its near unicity as the indicating a TOE.

Every correct investigation of localization by string theorists led to the
pointlike result. The safest calculation is that via the commutator of two
string fields. All these calculations led to one result: a pointlike localized
spacelike (graded) c-number commutator, whose explicit form still depends on
the choice of the internal part (the "vertically" acting oscillators) of the
smearing function \cite{Dim}. With other words the infinite mass/spin tower
spectrum is a general characteristic property of the theory, but the strength
with which these levels contribute to a particular point-localized wave
function or second quantized field analog can be manipulated with operators
acting between the levels.

String theorist in their correct calculations of (graded) commutators of
string fields of course do not obtain anything but a pointlike localization
\cite{Mar}\cite{Lowe}. But being members of a globalized community of string
theory they do not present their result in this form. Their subconcious desire
to serve the community course leads them to describe a conceptual chimera:
some sort of extended but at the same time hidden object, a kind of invisible
string of which only the c.m. point is sticks out. This shows to what extend
the critical power, which was keeping particle theory strong during healthy
times, was lost in the sociology of globalized communities and their guru like
leaders This phenomenon is one of the strangest I ever came across.

Remembering that the conquest of quantum theory is inexorably linked with a
clear exposition of quantum reality and localization in particular, one
wonders why string theory leads people to mystical regressions. The cited
papers constitute an interesting historical document for a time in which clear
calculations could not prevent their metaphoric interpretation. The tower of
Babel in particle theory is erected on the difference between computations and
prevailing ideology. It is of course important that the calculations are
correct, and it is not plausibe that the interpretation which fails to match
the calculation was distorted on purpose. The tower of Babel effect is rather
the result of the Zeitgeist in the service of a dominating TOE.

Perhaps the path into a self-defeating metaphoric world started already with
such innocent looking choice of notation which feigns string-like target space
localization as writing $X^{\mu}(\tau)$ for the current potentials $\Phi^{\mu
}(\tau)$ or even before by introducing the notion of target space (field
space) which has no place in QFT and is at best a metaphor for "arena of
action for inner symmetries". With the loss of conceptual knowledge about
local quantum physics, the idea of a stringlike target space localization may
have received a helping hand from an unlucky notation which could have
exacerbated already present misunderstandings.

String theory unlike QFT has no built-in operational way of introducing
interactions. Whereas the spacetime principles underlying QFT are strong
enough to not only determine the form of interactions consistent with the
locality principle but also to rigorously derive scattering theory, all these
ideas of deriving global properties from local principles are lost in a pure
S-matrix approach. Its principles of unitarity, Poincar\'{e} invariance and
possibly crossing are the only guides and every additionally imposed structure
has to justify itself a posteriori by its phenomenological success. Hence it
is not surprising that interactions are defined "by hand" via highlighting
certain operators which already played that role in the dual model.

Being deprived of large time asymptots which relates the S-matrix with a
Lagrangian via interpolating fields, string theorists simply define the lowest
order (tree approximation) of the string S-matrix by functional formulas which
are equivalent to the the Fubini-Virasoro exponential expressions. Already in
the setting of the dual model, attempts were made to find reasonably looking
recipes to imitate the loop corrections of QFT by adapting Feynman's rules for
world lines to world sheets. String theorists introduced computational recipes
in form of graphical descriptions in terms of rules for combining and
splitting tubes which are supposed to represent the world sheet traced out by
closed strings, but what does this mean for pointlike objects whose spacetime
string extension is metaphoric and not real? Whereas such recipes in QFT can
be shown to be a graphical illustration of operator relations, their quantum
meaning in string theory remain unclear. The characteristic feature of a
relation formula or idea in quantum theory is that it can be expressed in
terms of operators and states; if this is not the case it is not part of QT,
this was the leimotiv of Bohr and Heisenberg which led them to the notion of
"observables". Despite a search over more than 4 decades for an operator
formulation behind those recipes for perturbative string S-matrix amplitudes
by the best minds in the string community, no such quantum theoretical
formulation was ever found.

In this context it is interesting to remind oneself that Stuekelberg
discovered Feynman rules precisely in this graphical recipe form. In his
studies of macrocausality properties of an S-matrix he realized that, whereas
the spacelike macrocausality amounts to the cluster factorization of the
S-matrix, there was a finer macrocausality property for asymptotic timelike
separation. A 3$\rightarrow$3 particle scattering for example should contain
the possibility that first 2 particles interact in form of a 2-particle
scattering and afterwards the third particle enters the causal future of the
first process and meets and interacts with one of the outgoing particles. He
showed that the timelike trajectory between the two local scattering centers
corresponds to a propagator with (what later became known as) Feynman's
$\varepsilon$-prescription, expressing the fact that the second interaction
happened later. By assuming that interaction regions can be idealized as
pointlike vertices he obtained the Feynman rules. Of course nobody, including
himself, paid much attention to such an ad hoc recipe. The general acceptance
came only with the derivation in terms of operators and states which started
with Feynman and found its most concise expression in the work of Dyson.

In the string case not only is there no operator formulation for the world
sheet picture, such a formulation would create a clash with the pointlike
nature of the free string. There remains of course the possibility that an
infinite collection of pointlike fields offers a new kind of pointlike
interaction which has no counterpart in the standard setting of polynomial
(possibly infinite degree) interactions. But even if such a possibility
exists, any quantum interaction must allow a formulation beyond recipes and
prescriptions in terms of the quantum setting of operators and states.

String theory, either in its factual infinite component pointlike setting, or
its metaphoric guise of a "invisible string" is markedly different from
(finite component) QFT if it comes to the notion of \textit{degrees of
freedom}. QFT has more phase space degrees of freedom than QM; whereas in QM
there is a finite number of degrees of freedom in a finite phase space volume,
the cardinality in QFT is described by a mild form of infinity (the
compactness or nuclearity property of QFT \cite{Haag}). This is precisely what
guaranties the existence of thermal states at any temperature and the causal
shadow property which states that the algebra of a spacetime region equals
that of its causal completion \cite{Swie} (the quantum counterpart of the
Cauchy wave propagation). Both properties are lost in string theory. In view
of its importance for the problem of holographic relations of QFTs in
different dimensions this issue will reappear (last section) in a different context.

\section{Modular localization, the KMS condition and the crossing property}

\ In order to become aware of the significant conceptual differences between
the crossing property and duality it is necessary to have a profound
understanding of crossing. In the sequel I will for the first time present
some recent insight on this problem within the setting of modular localization
theory (appendix).

The important concept from modular theory which relates to the crossing
property is \textit{localization equivalence with respect to the wedge W
spacetime region\footnote{Although localization equivalence can be defined
between operator algebras which share the same Poincar\'{e} representation
theory in the same Hilbert space, only the wedge situation leads to the
crossing relation.} (}which will be denoted denote by\textit{ }$\overset
{W}{\sim})\ $\textit{ }between operators affiliated to ($\prec$) different
wedge algebras $\mathcal{A}(W)$ and $\mathcal{B}(W)$ which live in the same
Hilbert space and share the same positive energy representation of the
Poincar\'{e} group.%
\begin{equation}
B\overset{W}{\sim}A:B\Omega=A\Omega,~~B\prec\mathcal{B}(W),~A\prec
\mathcal{A}(W) \label{bi}%
\end{equation}
Since under such conditions modular theory identifies the dense subspaces
generated by applying the two wedge algebras\footnote{More precisely modular
theory identifies the range of the two algebras after closing it in the graph
norm of shared $\Delta^{\frac{1}{2}}$ which defines the same dense subspace.
This domain of Wightman fields is believed to include that subspace but the
range of those $B(f,..)$ which are l.e. in the expained sense is smaller (see
later).} to the vacuum, it brings about a one to one relation between
generally unbounded operators which does not respect the algebraic
multiplication structure. Hence the $\overset{W}{\sim}$ relation is a
\textit{bijection} between the individual operators affiliated to two
wedge-localized operator algebras which both live in the same Hilbert space
and share the same unitary representation of the Poincar\'{e} group, but may
be very nonlocal relative to each other. The situation which is relevant for
the derivation of crossing is that in which $\mathcal{B}(W)$ is the
wedge-localized algebra from an interacting net of local algebras which admits
a complete asymptotic interpretation and $\mathcal{A}(W)=\mathcal{A}_{in}(W)$
is the wedge algebra generated by its asymptotic incoming fields.

It is convenient for the following to introduce a flexible notation. If we
want to refer everything to the algebra $\mathcal{A}$ we will use the notation
$B_{\mathcal{A}}$ when we want to substitute a $B\in\mathcal{B}$ by its
bijectively related operator (\ref{bi}) $B_{\mathcal{A}}\in\mathcal{A};$
conversely we write $A_{\mathcal{B}}$ if our aim is the characterization of an
operator in the algebra $B$ which is bijectively related to $A\in\mathcal{A}.$
Returning to our situation of interest of an asymptotically complete theory
and $\mathcal{B}=\mathcal{B}(W),$ $\mathcal{A}=\mathcal{A}_{in}(W)$ we can
picture a $A_{\mathcal{B(}W\mathcal{)}}$ in a more concrete fashion: it is an
operator in $\mathcal{B}(W)~$whose creation component (the one involving only
creation operators $a^{\ast}(p)$) is identical to the creation component of
$A\in\mathcal{A}(W).$ The other components are uniquely determined by the
requirement that the operator belongs to a particular W-localized algebra.

The underlying idea resembles in some sense the algebraic notion of
\textit{relatively local fields} which led to the concept of Borchers
equivalence class \cite{STW}. But since there is no furthergoing algebraic
connection beyond the bijection between the operators of two local algebras
which only share the same localized states and the same representation of the
Poincar\'{e} group (and as a consequence, the same Reeh-Schlieder subspace
\cite{Haag}), the two algebras may be quite different in the algebraic sense
as exemplified by an interacting wedge algebra and its (via scattering theory)
associated asymptotic incoming algebra.

The existence of this bijection is a straightforward generalization of an
argument about modular theory in \cite{BBS}. In that work the interacting
representation of a wedge-localized one-particle state\footnote{Such operators
were called PFGs (vacuum-polarization-free-generators). They allow to
generalize the Jost-Schroer theorem (saying essentially that interacting
theories cannot have compact localizable PFGs) and play a crucial role in the
modular construction of factorizing models (see next section).} was
considered. Such vacuum polarization-free objects are not available in
interacting theories for compact localization region, in fact the wedge region
is in passing from compact to non-compact localized causally closed spacetime
regions the first for which such interacting one-particle generators exist. In
a more intuitive formulation: \textit{wedge regions lead to the best
compromise between particles and fields} in the presence of interactions. Only
algebras generated by free fields have vacuum polarization free generators for
\textit{any} localization region. Hence localization-caused vacuum
polarization clouds offer an autonomous criterion for the presence of interactions.

In a journal on foundations of physics it may be appropriate to mention that
these dense subspaces have attracted the attention of renown philosophical and
foundational motivated physicists. \cite{Cl-Hal}\cite{Ea-Rue}\cite{Fra}%
\cite{Rob}\cite{Summers}; in fact this has been a small window of intense
communication between physics and philosophy to which the critical remarks in
the introduction do not apply. In fact the existence of these subspaces was a
surprise for those who obtained her/his physical intuition from QM; they
constitute one of the most characteristic features of QFT. Although the
domains $domS$ only depend on the representation of the Poincar\'{e} group
(the mass/spin spectrum), the way how the different $S$ act on this domain
carries the informations about the interaction (appendix).

In the case the algebras generated by the cyclically acting fields are
identical $\mathcal{A}(W)=\mathcal{B}(W),$the bijection $\overset{W}{\sim}%
$\ leads back to the trivial relation $A=B.~$Hence it is a generalization of
the algebraic notion of local equivalence which is closely related to the
notion of the \textit{Borchers class} of relative local fields. Both concepts
are also related (but not identical) to \textit{weak locality~\cite{STW}}.

The bijection concept comes with a prize. If the operator $A(f)\prec
\mathcal{A}_{in}(W)$ \ is a $f~$-~smeared covariant field with $suppf\subset
W,~$having the standard Wightman domain properties, the existence of
$B^{\prime}s$ is paid for by unwieldy domain properties. Although acting on
the vacuum they do induce the same dense space of states; their domain
properties are weaker than those of smeared Wightman fields. Their generally
smaller domain is not translational invariant i.e. the translated domain of an
operator $B\prec\mathcal{B}(W)$ is outside $domB~$\cite{BBS}$.$ The
translation invariance of the domain (the \textit{temperateness} of $B$) would
imply $S=\mathbf{I}$ if $d>1+1,$ whereas in case d=1+1 the model has only
elastic scattering \cite{BBS}. This shows that modular theory does not only
reveal deep connections between spacetime geometry and the mathematics of
operator algebras, but also sheds new light on connections between domain
properties of unbounded operators and the presence of interactions.

For $S_{scat}=\mathbf{1}$ to occur it is enough if such a Poincar\'{e}
invariant dense domain exists for a particular $B(f)\prec\mathcal{B(W)}$ which
is in bijection with $A_{in}(f),$ supp$f\in W$
\begin{equation}
B(f)\Omega=A_{in}(f)\Omega
\end{equation}
i.e. a temperate B which \textbf{g}enerates a vacuum \textbf{p}olarization
\textbf{f}ree \textit{one particle state} (such a B is called a \textbf{PFG}%
\footnote{PFGs do not exist for causally complete subwedge regions unless the
theory is generated by a free field. (stronger than the triviality of
scattering). The wedge is the "smallest" causally closed region for which PFGs
exist, though generally only at the prize of nontranslational invariant
domains. Well behaved ("temperate") PFGs for Ws only exist in d=1+1.})
\cite{BBS}. The triviality of the scattering matrix $S_{scat}=1,$~and
therefore the equality of the Tomita operator $S_{Tomita}=S_{free}$ with that
of a free field follows (as long as one avoids low dimensions ($d=1+1$) and
and 3-dimensional models with plektonic statistic). The interesting question
to what extend this implies the absence of interaction in the stronger sense
of $\mathcal{B=A}_{in}$ will be commented on later.

The case of factorizing models, for which the S-matrix is nontrivial but has a
rather simple structure, will be presented in detail in the next section.

\bigskip The important relation which leads to the derivation of the crossing
property is \cite{BBS}%
\begin{align}
B\Omega &  =\Phi,\text{ }\Phi=A\Omega,~i.e.~\Phi\in domS_{\mathcal{A}},~B\in
B(W)\label{rep}\\
&  \curvearrowright B^{\ast}\Omega=S_{\mathcal{B}}\Phi\nonumber
\end{align}
This is a formula for the computation of the action of the conjugate of an
operator on the vacuum if the operator itself is unknown except that its
action on the vacuum should result in a state vector $\Phi$ which has no
direct relation to the $\mathcal{B}(W)$ algebra apart from its membership to
the space $domS_{\mathcal{A}}=domS_{\mathcal{B}}$. The theorem tells one how
to compute $B^{\ast}\Omega$ from these data. These prerequisites are always
met if the two algebras share the same representation of the Poincar\'{e}
group i.e. have the same mass/spin particle content.

The crossing relation in its simplest field theoretic formulation
(selfconjugate spinless fields, only incoming fields in the uncrossed
configuration) reads%
\begin{equation}
\left\langle B(A_{in}^{(1)})_{\mathcal{B}}A_{in}^{(2)}\right\rangle
=\left\langle A_{out}^{(2)}{}\Delta BA_{in}^{(1)}\right\rangle \label{orig}%
\end{equation}
It is important to note that $A_{in}^{(1)}$ on the left hand side appears as
its bijectively related counterpart $(A_{in}^{(1)})_{\mathcal{B}}$ which off
vacuum represents a different operator. $A_{in}^{(1)}$ and $A_{in}^{(2)}$ may
be products of W-smeared free fields%
\begin{equation}
A_{in}^{(1)}=:A_{in}(g_{1}),..A_{in}(g_{k}):,~~A_{in}^{(2)}=:A_{in}%
(f_{1})...A_{in}(f_{l}):
\end{equation}
The proof of this relation is reminiscent of the modular derivation of the KMS
relation, in fact if $A_{in}^{(1)}=1=(A_{in}^{(1)})_{\mathcal{B}}$ it reduces
to the above extension of the KMS property.

In the present case there are not only operators from different algebras to
start with, but the action of $S_{B}$ on $A_{in}^{(2)}$ brings a third algebra
into the game namely $\mathcal{A}_{out}(W).$ As in the case of KMS, the
presence of the unbounded analytically continued operator $\Delta$ leads
precisely to the same analytic properties as those found by Araki.

The formfactor crossing%
\begin{equation}
\left\langle B|p_{1}..p_{k}q_{1},.q_{l}\right\rangle _{in}=~_{out}%
^{a.c.}\left\langle -q_{1}..-q_{l}|B|p_{1}...p_{k}\right\rangle _{in}^{c.o}
\label{cro}%
\end{equation}
results from the previous field theoretic crossing if one takes instead the
over all Wick product $:(A_{in}^{(1)})_{\mathcal{B}}A_{in}^{(2)}:.$ But what
are the precise conditions under which the subsript ()$_{\mathcal{B}}$ can be
omitted? This will be explained below. The W-localized wave functions which
were still present in the field theoretic crossing (\ref{orig}) have been
removed in (\ref{cro}) as the result of their on-shell denseness so that the
crossing identity only involves momenta\footnote{In the case of non
selfconjugate particles the $q$-momenta refer to antiparticles and it would be
better to use the notation $\bar{q}.$}. The notation is as follows, the $a.c.$
refers to the analytic continuation from the positive mass shell to the
backward shell (using the momentum space analyticity of wedge localized
mass-shell reduced test functions) and the $c.o.$ indicates the omission of
contractions between the $p^{\prime}s$ and $q^{\prime}s$ which reflects the
fact that the l+k particle state on the right hand results from a Wick-ordered
product of in-fields; since there are no contractions between in-particle on
the right hand side, there can be none after crossing to the left hand side
either. The negative momenta $-q$ are a result of the combined action of
$S^{\ast}=\Delta^{\frac{1}{2}}J$ where $\Delta^{\frac{1}{2}}$ has the
geometric interpretation of an imaginary $\pi$ Lorentz rotation $\Delta
^{\frac{1}{2}}=e^{-\pi iK};$ these particles are outgoing and they would be
antiparticles in case the particles carry a superselected charge (the fields
are not selfadjoint).

For the proof one uses the formula (\ref{rep}) (first line, second equation)%

\begin{align}
\left\langle B(A_{in}^{(1)})_{B}A_{in}^{(2)}\right\rangle  &  =((B(A_{in}%
^{(1)})_{\mathcal{B}})^{\ast}\Omega,A_{in}^{(2)}\Omega)=(S_{\mathcal{B}%
}(BA_{in}^{(1)})\Omega,S_{\mathcal{B}}A_{out}^{(2)\ast}\Omega)=\label{rel}\\
&  =\left\langle A_{out}^{(2)}\Delta BA_{in}^{(1)}\right\rangle \nonumber
\end{align}
where in the the last line the antilinearity of $S_{\mathcal{B}}$ as well as
the relation $S^{\ast}S=\Delta$ was used. \ Apart from the involvment of
different algebras, the derivation of the crossing relation resembles strongly
the modular derivation of the KMS property of localized algebras and may be
seen as a \textit{generalization of the KMS setting}. In fact the relation
(\ref{rel}) is a KMS relation in the presence of two different wedge-localized
algebras$~\mathcal{B}(W),\mathcal{A}(W)$ which share the same representation
of the modular group (the Lorentz boost). In the case at hand, the joint
modular group results from the sharing of the same Poincar\'{e} group
representation between the interacting theory and its asymptotes. It is a
special case of an \textit{extended KMS relation for two algebras} which are
standard with respect to the same vector state and have the same modular group
but different modular reflections $J_{\mathcal{A}}\neq J_{\mathcal{B}}$
\begin{equation}
\left\langle BA\right\rangle =\left\langle ((A^{\ast})_{\mathcal{B}})^{\ast
}\Delta B\right\rangle ,~(A_{\mathcal{B}})^{\ast}\Omega\equiv S_{\mathcal{B}%
}A\Omega
\end{equation}
with $S_{\mathcal{B}}$ being the modular Tomita operator for the algebra
$\mathcal{B}(W)$\footnote{The explicit computation of the action of a modular
$S$- operator on a state generated by a $\Delta$-related operator algebra is
generally a difficult problem.}, the two-algebra generalization of the KMS
situation evidently reduces to the one-algebra case for $\mathcal{B}%
(W)=\mathcal{A}(W)~$and hence$~A_{\mathcal{B}}=A_{\mathcal{A}}\equiv A.$ In
the interacting case only the creation components coalesce; the statement that
all other (model-dependent) contribution to the formfactor contain momentum
space delta functions and are removed under the c.o operation will be deferred
to a separate publication\footnote{J. Mund and B. Schroer: "A generalized KMS
condition and its relation to the crossing property" in preparation.}. This
two-algebra extension of KMS offers a wealth of new applications whose
presentation would go beyond the theme of this essay.

The formfactor relation which follows from (\ref{rel}) has the follwing form
in momentum rapidity space%
\begin{equation}
\left\langle B\left(  a^{\ast}(\theta_{1})..a^{\ast}(\theta_{k})\right)
_{\mathcal{B}}a^{\ast}(\vartheta_{1})..a^{\ast}(\vartheta_{l})\right\rangle
=\left\langle a(\vartheta_{1})..a(\vartheta_{l})Ba^{\ast}(\theta_{1}%
)..a^{\ast}(\theta_{k})\right\rangle
\end{equation}
For simplicity of notation we specialized to two spacetime dimensions
(omission of transverse momenta components) and assumed that the particles are
selfconjugate. Wick-products of Bose fields are symmetric in their arguments
and this symmetry property is inherited by the (..)$_{\mathcal{B}}.$ It is
custumary in formfactor constructions to get rid of this redundancy and make
the convention that n-particle states are always $\theta-$ordered i.e.
\begin{equation}
Ta^{\ast}(\theta_{1})..a^{\ast}(\theta_{k})=a^{\ast}(\theta_{i_{1}})..a^{\ast
}(\theta_{i_{k}})~~if\text{ }\theta_{i_{1}}>\theta_{i_{2}}>...>\theta_{i_{k}}%
\end{equation}
But since $a^{\ast}(\theta)_{\mathcal{B}}\equiv b^{\ast}(\theta)$ is generally
a very complicated operator whose only simplicity consists in that it defines
a wedge-localized PFG (i.e. its application to the vacuum created a
one-particle state without vacuum polarization), the problem is to define a
symmetric product which agrees for ordered arguments with the operator product
and which does not require any special commutation relations as those which
characterize free fields. The only answer is the \textit{theta-ordered
product\footnote{Note that not even in the standard perturbative setting the
time ordered product is simply the time-ordering of the unordered product in
any naive sense.}}. One then obtains the desired composition rule%
\begin{align}
(Tb^{\ast}(\theta_{1})..b^{\ast}(\theta_{k}))a^{\ast}(\theta_{1})..a^{\ast
}(\theta_{k})\Omega &  =a^{\ast}(\theta_{1})..a^{\ast}(\theta_{k})a^{\ast
}(\vartheta_{1})..a^{\ast}(\vartheta_{l})\Omega\\
if\text{ }\theta_{i} &  >\vartheta_{j}~\forall~i,j\text{ }\nonumber
\end{align}
i.e. the interacting ($a^{\ast}(\theta_{1})..a^{\ast}(\theta_{k}%
))_{\mathcal{B}}\equiv$ $Tb^{\ast}(\theta_{1})..b^{\ast}(\theta_{k})$
operators in our bijection act as products of interaction-free operators only
in special orderings between T-ordered clusters. Only in the well-studied case
of factorizing models the $b$-operators have simple properties, namely they
have a translation invariant domain and obey Zamolodchikov-Faddeev commutation
relations (see next section). In that case one also knows the analytic
continuation properties under exchange of rapidities which are not part of the
crossing relation (which only refers to the cyclic KMS permutations). More on
the properties of these bijectively related operators will be contained in
forthcoming joint work \cite{M-S}.

Since the analyticity properties result from the domain properties of
$\Delta,$ it is helpful to remind the readers of the standard analytic KMS
properties as Araki \cite{Araki} first established them.

\begin{definition}
Let $\mathfrak{C}$ be a C$^{\ast}$algebra on which $\alpha_{t}$ acts as a one
parameter automorphism group. Then $\omega$ is called a KMS state with respect
to $\alpha_{t}$ at temperature $\beta>0$ if for each pair of operators $A,B$
$\in~\mathfrak{C}$ there exists a function $F_{A,B}(z),$ analytic on the open
strip $\left\{  z\in C,~0<\operatorname{Im}z<\beta\right\}  ,$ continuous and
bounded on its closure, such that%
\begin{equation}
F_{A,B}(t)=\omega_{\beta}(Aa_{t}(B)),~~F_{A,B}(t+i\beta)=\omega_{\beta}%
(a_{t}(B)A) \label{KMS}%
\end{equation}

\end{definition}

Araki showed that the n-point correlation functions in a KMS state are
boundary values of analytic functions in the strip $C_{\beta,<}^{(n)}$ given
by
\begin{align}
&  \omega_{\beta}(\alpha_{t_{1}}(B_{1})....\alpha_{t_{n}}(B_{n}))=\lim
_{\operatorname{Im}z\rightarrow0}\omega_{\beta}(\alpha_{z_{1}}(B_{1}%
)....\alpha_{z_{n}}(B_{n}))\label{Araki}\\
&  C_{\beta,<}^{(n)}:0<\operatorname{Im}z_{1}<.....<\operatorname{Im}%
z_{n}<\beta\nonumber
\end{align}
\ and $\omega_{\beta}$ exists under rather general conditions for all
$\beta>0.$

There are similar analytic properties of KMS states which come with only half
the strip region \cite{Araki}%
\begin{align}
&  \alpha_{t_{1}}(B_{1})....\alpha_{t_{n}}(B_{n})\Omega_{KMS}=\lim
_{\operatorname{Im}z\rightarrow0}\alpha_{z_{1}}(B_{1})....\alpha_{z_{n}}%
(B_{n})\Omega_{KMS}\\
&  C_{\beta/2,<}^{(n)}:0<\operatorname{Im}z_{1}<.....<\operatorname{Im}%
z_{n}<\beta/2\nonumber
\end{align}
Note that there is no statement on whether different orderings can be related
by analytic continuation; in general this is not possible. In the case of
Wightman functions however this follows from spacelike (graded) commutativity,
and for the so called temperate PFGs of d=1+1 factorizing theories this is a
consequence of the Zamolodchikov-Faddeev commutation relations for generators
of wedge localized algebras (next section).

In the case at hand the thermal aspect does not arise in the standard way i.e.
by subjecting a global algebra of QFT to a heat bath which converts its ground
state into a KMS state. It rather originates by restricting a global vacuum
state to a wedge-localized subalgebra. With the conventions from modular
theory we have%
\begin{align}
\Delta^{it}  &  =~e^{-2\pi itK},~K=generator~of\ W-Lorentz~boost\\
\beta_{\operatorname{mod}}  &  =-1~corresponds~to~\beta=2\pi\nonumber
\end{align}
Whereas the Hamiltonian and the temperature $kT=\beta^{-1}$ are dimensionful
quantities, the modular temperature and the modular Hamiltonian are
dimensionless since they arise in a geometric context.

This raises a very fundamental question: is the KMS analytic aspect of
crossing with real thermal physics only a parallelism in the mathematical
formalism or does it extend to the physical content.

The question how the basic quantities of a heat bath situation, as energy and
entropy, are related to their counterparts arising from localization is a
fundamental problem of quantum theory, in view of its astrophysical
applications perhaps the most fundamental problem of our times. Although its
understanding does not contribute anything directly to the crossing property,
some general comments on such a pivotal problem are in order. Both problems
are related to KMS states on the same algebra namely the hyperfinite type
III$_{1}~$factor algebra. In \cite{BMS} the reader finds rather tight
arguments that the thermodynamic infinite volume limit of a heat bath system
corresponds to a certain "funnel" approximation of a localized algebra by a
family of slightly larger algebras defined in terms of "the split property" in
such a way that the divergent volume limit for the entropy can be placed in
direct correspondence with a logarithmically corrected divergent area law.

It is to be expected that such a method, even if ingeniously applied as in
\cite{BEG}, is too bulky for a general solution of the crossing problem, in
particular in view of the fact that it does not point to the relevant physical
setting (KMS from localization). Its exploration came to an end already in the
70s after K\"{a}ll\'{e}n and Wightman tried for many years in vain to derive a
general representation of a 3-point function on the basis of computations of
natural muti-variable analyticity domains.

Historically thermal properties of localization entered QFT through the
Hawking radiation of quantum matter behind an event horizon. For some time
this was thought of as a separate issue of QFT in curved spacetime. But the
main difference between event horizons in curved spacetime and causal horizons
in Minkowski spacetime QFT is that the former are objective locations given by
the external metric, whereas the latter are Gedanken-constructs whose physical
realization depends on non-inertial observers (viz. the Unruh effect). The
fleeting existence (i.e. not experimentally realizable) of causal horizons
does not at all mean that they are unimportant for a structural comprehension
of QFT. The fact that the insufficiently understood crossing property of
particle physics reveals it full physical significance in the setting of
thermal manifestations of modular localization confirms this. This confluence
of particle physics concepts with concepts coming from black hole physics is a
very exciting process of ongoing conceptual unification which promises to
bring a wealth of new insights.

There are interesting structural consequences of the crossing property, e.g.
the Aks theorem \cite{Aks} stating that d%
$>$%
1+1 quantum fields cannot lead to elastic scattering without the presence of
inelastic scattering processes. The factorizing models in d=1+1 are an
exceptional case; such models carry the full infinite vacuum polarization, but
its S-matrices are certain combinatorial products of two-particle
$S_{scat}^{(2)}(\theta_{1}-\theta_{2}).$ Another expected consequence of
localization equivalence and crossing is that $S_{scat}=1$ implies that the
theory is that of a free field\footnote{I am indepted to Jens Mund who
informed me about a forthcoming paper by him on this generalization of the
Jost-Schroer theorem.}, but the arguments given in \cite{inverse} can
presently only be made rigorous for factorizing models.

Crossing is a consequence of the specific field theoretic (modular)
localization and not a general property of relativistic QT. There exists a
relativistic particle quantum mechanics, the DPI (direct particle interaction
theory) \cite{interface} which is based on the non-covariant
Born-Newton-Wigner localization \cite{N-W} resulting from the spectral
decomposition of the selfadjoint position operator. The DPI Hilbert space
carries an interacting multiparticle representation of the Poincar\'{e} group
which fulfills the cluster factorization property. However it contains no
covariantly localized objects at finite times, the only covariant object is
the (global) S matrix which is invariant and has the cluster decomposition
property for spacelike directions (macrocausality). In fact it fulfills all
properties which one is able to formulate in terms of particles
\cite{interface}.

On the other hand the properties presented in this section need the causal
relativistic localization which, although leading to important consequences
for particle scattering (as crossing), cannot be understood in a \textit{pure}
particle setting. The velocity of light in DPI setting, similar to the
velocity of acoustic waves, comes about through quantum mechanical state
averaging at large times; it refers to the center of a wave packet whereas in
QFT it is a microscopic property of the observable algebra which is not
related to the c.m. movements of wave packets. The good news is however that
in QFT the \textit{BNW localization becomes asymptotically covariant} and thus
consistent with modular localization. In particular the asymptotic
interpretation of QFT inherits the BNW probability without which one could not
obtain invariant scattering cross section.

\section{An exceptional case of localization equivalence: d=1+1 factorizing
models}

In the generic setting of formfactor crossing there is no property by which
one can \textit{interchange} the position of the rapidities by the process of
analytic continuation\footnote{The particle statistics (Bosons, Fermion) is
used to bring the rapidities into the natural order. The n! natural orders are
generally belonging to n! analytic functions which are not analytic
continuations of each other.}. For the d=1+1 factorizing models this is
however possible i.e. there is only one analytic masterfunction which relates
\textit{all} rapidity orderings. This additional analytic structure makes it
possible to use the extended analytic setting as the start of a classification
and explicit construction of models through the S-matrix and the formfactors
of its fields: the bootstrap-formfactor project \cite{Ba-Ka}.

From the viewpoint of modular localization based construction favored in the
present paper, these properties turn into powerful tools of model
constructions. These models are distinguished by the fact that their wedge
algebra contains what has been referred to as "temperate PFGs" (vacuum
\textbf{p}olarization-\textbf{f}ree \textbf{g}enerators) \cite{BBS}. PFGs are
operators operators which applied to the vacuum have translation invariant
domains and, as a consequence, well behaved Fourier transforms. With other
words the d=1+1 $B$-fields which are bijectively related via their shared
wedge localized state space to the wedge-localized state space generated by
incoming/outgoing free fields have now translational invariant domains (i.e.
are temperate). It turns out that all so-called factorizing models
\cite{Ba-Ka} are in this class and it appears that temperate PFG always lead
to factorizing models. The covariant domain properties result in the existence
of a wedge-independent on-shell Fourier transformation leading to a free field
like representation \cite{BBS} which for the simplest family of models (the
Sinh-Gordon model) \cite{AOP}\cite{Lech1} are of the form%

\begin{align}
&  \Phi(x)=\frac{1}{(2\pi)^{\frac{3}{2}}}\int\left(  Z^{\ast}(\theta
)e^{ipx}+h.c.\right)  \frac{d\theta}{2},~~p=m(ch\theta,sh\theta)\label{ZF}\\
&  Z(\theta_{1})Z(\theta_{2})=s(\theta_{1}-\theta_{2})Z(\theta_{2}%
)Z(\theta_{1}),~Z(\theta_{1})Z^{\ast}(\theta_{2})=s(\theta_{1}-\theta
_{2})Z^{\ast}(\theta_{2})Z(\theta_{2})+\delta(\theta_{1}-\theta_{2})\nonumber
\end{align}
where it is convenient to use the mass shell rapidity instead of the mass
shell momentum. Here $s$ is the two-particle scattering function of the
Sinh-Gordon model; in the general case of factorizing models the $Z$-operators
are multi-component creation/annihilation operators and the scattering
function becomes a scattering matrix.

The Z-commutation relations are a special special case of the
Zamolodchikov-Faddeev algebra structure, but in contrast to their original use
as pure algebraic calculational devices, the Z's in the present wedge
localized approach have a spacetime interpretation. Although the affiliated
field $\Phi$ for $s\neq1$ lacks pointlike localization, it can be shown to be
at least wedge-like localized \cite{Sch}. When these properties became clear
during the 90s \cite{AOP}, the guiding idea that the larger the spacetime
region one has at one's disposal, the easier the control of vacuum fluctuation
andr to find simple generators of the localized algebra\footnote{The knowledge
about the system however decreases with increaing localization size.}.
Pointlike local fields are of course generators for algebras localized in
arbitrary spacetime regions, but in the new constructive approach they appear
at the end as the cherry on the cake.

The best localization region below the full algebra associated with the
Minkowski spacetime which still admits a particle structure is the
(noncompact) wedge region. this algebra is the "smallest" which contains for
the first time PFG operators i.e. operators which once applied to the vacuum
behave like a free field but have a complicated action on other states; i.e.
although far more involved than free fields, in their application to the
vacuum they behave precisely like a free field. This was the beginning of a
new construction principle which I applied to factorizing models \cite{Sch}
before Gandalf Lechner \cite{Lech1} used it to proof the first existence
theorem of the strictly renormalizable (short distance singularities involve
powers worse than those of free fields), but not superrenormalizable models.
The Fourier transforms of the wedge generating fields were the Z-F operators
of the above form.

In the standard terminology $\Phi$ is a \textit{nonlocal} on-mass-shell
covariant field, but an application of modular theory shows that it is far
from being completely nonlocal since it is wedge localized \cite{Lech1} in the
sense that smeared with W-supported test functions $\Phi(f)\prec
\mathcal{B}(W).$ Contrary to free fields for which the localization is
entirely governed by the support of the test function, the use of compact
localized test function inside W does not improve the situation.

The possibility of "localizing in momentum space" in d=1+1 i.e. to work with
operators $Z(\theta)$ (\ref{ZF}) with Wightman-like domain simplifies the
discussion and permits to arrive at more detailed results than the crossing of
the previous section where algebraic properties of the operators $B(f,..),$
which are comparable to those of the temperate PFG generators $Z,$ are not available.

There exists a very simple-minded almost kinematical argument why in d=1+1 the
temperateness of wedge localized PFGs does not exclude interactions. It so
happens that the two-dimensional energy-momentum conserving delta function
coalesces with the tensor product of two particle mass shell delta functions
which appear in the inner product of a two-particle state. This has as a
consequence that the \textit{cluster factorization argument} for the S-matrix
cannot distinguish between an elastic $S^{(2)}$ and a trivial $S^{(2)}%
=\mathbf{1}$ i.e. clustering in d=1+1 cannot remove a two particle interaction
and arrive at a trivial scattering amplitude. In this sense the models stay
close to non-interacting situations. Nevertheless the \textit{off-shell
structure of these models is surprisingly rich}, in particular they possess
the full vacuum polarization struture for compact spacetime localization,
although they have no on-shell particle creation through scattering. Their
mathematical and conceptual structure has been the object of many studies and
they continue to play the role of a theoretical laboratory in which quantum
field theoretical ideas can be tested and studied under full mathematical control.

The states obtained by the iterative application of the $Z$ have a very simple
structure%
\begin{align}
&  TZ^{+}(\theta_{1})....Z^{+}(\theta_{n})\Omega=a_{in}^{\ast}(\theta
_{1})...a_{in}^{\ast}(\theta_{n})\Omega\\
&  \bar{T}Z^{+}(\theta_{1})....Z^{+}(\theta_{n})=a_{out}^{\ast}(\theta
_{1})...a_{out}^{\ast}(\theta_{n})\Omega\nonumber
\end{align}
where $T$ is the $\theta$-ordering (same symbol as for time-ordering) and
$\bar{T}$ denotes the opposite ordering and the right hand side only involves
symmetric Bose operators. The analytic properties of the vacuum polarization
component for a fixed order%
\begin{equation}
F(\mathcal{O},\theta_{1}...\theta_{n})\equiv\left\langle 0\right.
|\mathcal{O}\left\vert Z^{\ast}(\theta_{1})...Z^{\ast}(\theta_{n}%
)\right\rangle ^{in},\text{ ~~}\theta_{1}>...>\theta_{n} \label{vac}%
\end{equation}
are those expected from the the previous section. But now, as mentioned
before, the analytic properties go beyond those coming from the cyclic KMS
property since the $Z$ commutation relations also encode what happens when the
order is analytically interchanged. This is similar to the extension of the
primitive tube domain of Wightman functions by the use of locality. On should
not confuse this commutation with (graded) bosonic statistics. The latter has
been already absorbed by encoding states which coincide after applying
particle statistics into one ordered master-state $\theta_{i_{1}}%
>..>\theta_{i_{n}\text{ }}$written as $\left\vert Z^{\ast}(\theta_{i_{1}%
})...Z^{\ast}(\theta_{i_{n}})\right\rangle $; it couples the $\theta$-order
with the operator order in products whereas the analytic change of order is
dynamic and extends Watson's observation that the boundary values of the
two-particle formfactor in the elastic region are determined by the elastic
part of the scattering matrix \cite{Ba-Ka}. Without this analytic interchange
it is not possible to understand the algebraic aspects of the work on the
bootstrap formfactor construction.

The knowledge about commutation properties is not availabe in the general
case; in the derivation crossing in the previous section we only used the
extended Araki KMS analyticity. Crossing does not tell anything about an
analytic exchange of two $\theta^{\prime}s,$ i.e. the analyticity which
permits to change the order of rapidities comes from the algebraic commutation
structure of the $Z$ generators. The crucial property which permits the
explicit computation of formfactors of fields is this analytic exchange of
rapidities (not to be confused with the exchange property due to particle
statistics) in which the factorizing S-matrix shows up. The Z-F commutation
relations result from the algebraization of this analytic structure. Its
higher dimensional generalization is the so called Watson theorem: the
difference between the upper and lower branch of the elastic scattering cut of
the two-particle formfactor is given by the elastic part of the S-matrix. The
introduction of rapidities "unfold" this cut, but since in non-factorizing
theories there exist all the higher inelastic cuts, a uniformization in terms
of rapidities which leads to a meromorphic function in the plane is not
possible. Hence the constructive power of factorizing models does not only
come from the general crossing property but rather results also from the
powerful analytic exchange property in conjunction with crossing.

The factorizing models confirm again that crossing has no conceptual relation
to duality. One-particle bound states which are poles in scattering processes
have no special place in crossing; models without or with boundstates fulfill
crossing and in case there are bound states present, they are mixed via
crossing with the scattering continuum in a complicated way. Even in
perturbative crossing the one-particle direct or exchange contributions do not
play any special role, there is no crossing in which only one-particle states
participate. 40 years of research on S-matrix based particle theory (duality,
ST) have been founded on misunderstandings of the crossing property.
Mathematically all the dual model constructions are synonymous with Mellin
transforms of conformal QFTs, a topic which physically could not be more
removed from crossing. So the discovery of dual models is nothing else than
the discovery of the infinite pole-structure of the Mellin transform of
conformal QFTs which in turn reflects the properties of the global operator
expansions \cite{Mack}. It is hard to imagine anything further from the
quantum field theoretical collision theory.

The full analytic setting of the so-called bootstrap-formfactor program (which
resulted from a correct understanding of crossing) was already formulated at
the late 70s \cite{Kar}; since that time there has been a steady stream of
novel models and new insights based on the analytic properties of their
formfactors \cite{Ba-Ka}. In all cases the calculated formfactors were not
only meromorphic functions in the multi-strip regions (where their poles have
a direct interpretation in terms of bound states), but they were even
meromorphic in the full complex $\theta$-plane (the infinitely many different
sheets in the Mandelstam s-t variables).

The conceptual basis of this approach received a significant boost when it was
observed that the analytic rules for the construction of formfactors permit a
formal algebraic encoding. What was first introduced as a trick without any
apparent intrinsic physical meaning \cite{Za} in the 90s acquired the
spacetime meaning of being closely related to wedge localization \cite{AOP}
which finally led to the first existence proof for factorizing models
\cite{Lech1}.

The interesting problem is to find an higher dimensional counterpart of these
observation. In the present context one certainly does not expect a simple
analog \ of $Z^{\#}$ operators which relate the different $\theta~$orderings
in the sense that the connection between the different $\theta$-orders in the
vacuum polarization formfactor (\ref{vac}) can be encoded into the operator
positions. The difficulty is that outside the temperate setting there is no
known mechanism by which one could get to an analytic exchange of two repidities.

To look for an algebraic interpretation of analytic continuation in terms of
an auxiliary QFT is not so absurd as it appears at first sight. The analogy
with Wightman theory is worth exploring. Wightman functions are distributions
whose primitive analytic properties come from the energy positivity. The
analytic tube regions for different spacetime orderings are related by the
algebraic properties of covariance and local commutativity. This gets quite
complicaed in case of d=1+2 braid group commutation structures where the
analytic continuation leads to multivalued functions. The formfactors in
factorizing theories are also multi-valued in the Mandelstam variables and by
rewriting this in the temperate case into the uniformizing $\theta$ variables
one finds an algebraic structure. The crucial question is whether the
analyticity properties of formfactors in the general case also permits to
encode the analytic change of $\theta$-orders into the position of generalized
Z-operators; such a property would go beyond crossing and could play an
important role in nonperturbative model constructions beyond factorization.

There are many more d=1+1 unitary elastic S-matrices satisfying crossing than
there are pointlike Lagrangian couplings i.e. most of the existing factorizing
models do not have a Lagrangian name. There is no reason to believe that this
is in any way different in higher dimensions, so there is a strong suggestion
that even outside factorizing models the Lagrangian formalism only covers a
tiny area.

As often with physical ideas, the best insight into their inner workings may
have little resemblance with the history which led to their discovery. Indeed
the original observation leading eventually to factorizing models had little
to do with what was presented in this section, in fact it was not even related
with factorizing S-matrices but rather with integrable looking quasiclassical
mass spectra of certain field theories (notable Sine-Gordon). In analogy to
integrable systems of QM as the hydrogen atom, it was natural to look for
higher conservation laws. But historically the first hints came from mass
shell restriction of perturbative correlation functions leading to scattering
amplitudes which were expected to show the absence of on-shell creation as an
indication of their integrability \footnote{As a curiosity I remember how one
of my Ph.D students (Bernd Berg) in the beginning of the 70s demonstrated such
statements numerically on one of the old Hewlett-Packard pocket calculators.}.

From such confidence-building calculations sprung the first suspicion that
behind these observation there was the S-matrix bootstrap, but this time
without the old ideological bombast \cite{Weiss}\cite{KTTW}. The first
structural arguments pointing into the direction of the S-matrix bootstrap
approach set off a frenzy of model classifications and construction according
to the bootstrap S-matrix program. It soon became part of a new
\textit{bootstrap-formfactor approach to factorizing models }(for more on the
history see \cite{Swie}).

\section{Resum\'{e}, some personal observations and a somewhat downbeat
outlook}

The era of post renormalization QFT began at the end of the 50s with a return
of the incompletely understood age-old \textit{particle-field problem}. The
formulation of the LSZ scattering theory and its rigorous derivation by Haag,
Ruelle and Hepp are important landmarks in this conquest. Another more recent
important step is the partial resolution of the apparent contradiction between
the noncovariant Born-Newton-Wigner localization, which brings the
indispensable probabilistic concept of QM\footnote{Born \cite{Born} introduced
this probability concept first in the setting of scattering theory (the Born
approximation for the cross section) before it was extended to x-space wave
functions.} into QFT, and the modular localization, which is intrinsic to QFT
but does not lead to the probability of finding a particle in a specified
spacetime region \cite{interface}. It is deeply satisfying that in the large
time scattering limit both localizations match; hence in particular the
noncovariant BNW localization becomes covariant\footnote{In the literature one
sometimes encounters an\ "effective" version stating that covariance is
attained for distances beyond the Compton wave length..} and the modular
localization becomes consistent with a probability concept which in turn is
the prerequisite for an invariance S-matrix and the probabilistic
interpretation of the associated cross sections.

This large time asymptotic coexistence between particles and fields or their
generated localized operator algebras is crucial for our understanding of QFT
and the crossing property is the (perhaps most subtle) manifestation of the
particle-field relation.

The first successful test of scattering theory consisted in the derivation of
the experimentally verified Kramers-Kronig dispersion relations from analytic
properties of field theoretic locality. This was important for strengthening
the confidence in the locality and spectrum principes of QFT.

It was in this context that the crossing relation arose in form of the
existence of an analytic masterfunction which connects different processes
with different distribution between incoming and outgoing particles. This was
a crossing identity in which the crossed in/out particles were in an
unphysical position. One still needed analytic continuation properties which
the LSZ scattering formalism by its own theory did not provide. For certain
scattering configurations this analytic argument was supplied in \cite{BEG}.
In the S-matrix bootstrap approach the crossing analyticity was simply assumed
under the heading "maximal analyticity", it was treated as a basic postulate
together with the other physical principles as Poincar\'{e} invariance and
unitarity. This way of looking at a problem by elevating a mathematical
property as analyticity to be on par with physical properties foreclosed the
chance to understand crossing in terms of localization and ensuing thermal KMS
properties; in particular the KMS-like cyclic permutation property (\ref{rel})
of scattering amplitudes and formfactors remained unnoticed.

Historically the next step was the successful use of the crossing relations
within the bootsstrap-formfactor program \cite{Ka-Wei} for factorizing models.
This did not involve a structural understanding; rather crossing was one of
the assumptions in the constrution of these models and the fact that at the
end one had constructed a nontrivial model meant that crossing is really a
property of this particular class of models. No connection to modular
localization properties and their thermal manifestations was noticed. This
changed with the realization that behind the Zamolodchikov-Faddeev algebraic
reformulation of factorizing models there are nonlocal wedge-localized
generators \cite{Sch}. Only then the construction in terms of recipes for
formfactors finally became a classification and construction of factorizing
models according to the underlying principles of QFT without the inference of
additional recipes \cite{Lech1}.

The derivation of crossing for formfactors presented in this paper is
according to my best knowledge the first one outside the narrow setting of
two-dimensional factorizing models. Since the context of this paper is a
rather broad one, a more detailed specific account of crossing from modular
localization theory and implications thereof will be given in a separate
publication \cite{M-S}.

The history of crossing shows also that an early flare-up of ideas, before
their conceptual-mathematical understanding is available, may under certain
sociological conditions cause disarray\footnote{Usually premature observations
disappear and return often in a different context when the understanding of
their conceptual-mathematical struture is in place \cite{Jor}.}. The dual
model and string theory and with it that strange idea of a millennium TOE
would not have come about without a certain amount of conceptual confusion. As
we know nowadays the properties of Mellin transform of conformal QFTs are
synonymous with dual models, including those first discovered by Veneziano and
others; for their construction one does not have to know anything about the
crossing property but only how to construct functions with a certain pole
structure in different Mellin variables. The pedestrian construction which
used properties of Gamma function in a very clever way was physically
interpreted as a one-particle approximations of the conjectured Mandelstam
representation for scattering amplitudes. It led to the belief that one had
discovered a deep and mysterious new area of particle physics outside of QFT,
whereas in reality it was the entrance into a physical no man's land.

The string theoretic extension of the dual model aggravated the problem of its
conceptual positioning, in particular since its pointlike localized nature was
overlooked as the result of confounding the presence of oscillators of a
quantum mechanical string with the presence of a string localized in
spacetime. The decisive factor which cemented this confusion through all those
decades up to the present was however the sociological impact of the
enthusiastic support by renown members of the physics community. Who will deny
the impact of statements about string theory as "a present of the 21st century
to the 20th", "there is no other game in town" or the citation of Churchills
famous die-hard slogan "never, never,...never give up" is living in an ivory tower.

At this point the difference to particle physics before the 80s becomes clear:
the fragile equilibrium between the innovative and speculative side of
particle physics and the critical counterweight had broken down\footnote{The
first version of the present paper was uploaded to arXiev:hep-th when a
moderator placed it to the general physics setion with a built-in barrier to
prevent any crosslisting of the paper. There is no more fitting description of
the present sociological state of particle theory, any commentary about this
episode is superfluous.}. The historically grown pre-electronic basic
knowledge about QFT appears now, in the presence of a millennium TOE
increasingly irrelevant. This is accompanied by a growing split between
applied QFT, where the main aim is to find computational recipes about a
subject which is thought of as having been basically understood, and
foundational research in LQP which is expected to lead to profound structural
discoveries by following the inner logic of the theory but often at the prize
of loosing contact with the actual reality of particle physics. There is
hardly any cross fertilization; the one side fails to penetrate the
conceptual-mathematical barrier to comprehend new structural insights into QFT
(and often thinks it is not even worth a try), whereas the other side has
distanced itself so much from the phenomena that even when one of their
findings can be connected to observational particle physics, it would probably
go unnoticed.

Speculative proposals with little conceptual support but popular appeal were
made at all times; particle theory is by its own nature a highly speculative
science where it is sometimes necessary to take a dive into the "blue yonder"
of the unknown. What was however different during the decades of dominance of
string theory is that the critical counterweight, which had quite a tradition
on the old continent, was not available after the 70s when it was most needed.
The leading figures in mathematical physics and (algebraic) quantum field
theory who had the conceptual insight to play this indispensible critical role
unfortunately did not enter the fray, and thus the old "Streitkultur" was
lost. In the beginning of this disengagement the phenomenological proposals of
Regge-trajectories were far removed from any structure which one could relate
with known principles of relativistic quantum theory; but when the sudden
transition to a fundamental TOE took place\footnote{The begin of modern string
theory has a date, it is the week in Paris in 1974 when Scherk and Schwarz
\cite{S-S} wrote up their famous paper. Underlining the rapidity of change one
may call it the Bartholomew-like massacre of the old string theory which
started with phenomenology of Regge trajectories.}, the uncritical acceptance
of the new string theory as a TOE happened with such a speed that a critical
discourse was hardly possible. The string protagonists occupied research and
university positions within a short time. Often their main credentials were
that they are working on the allegedly most important millennium theory. After
some of the leading High Energy laboratories began to hire string theorists,
it was a matter of national and scientific pride to have a representative of
string theory as a kind of signboard of participation in the new millennium project.

In order to avoid misunderstandings, the derailment of parts of particle
physics caused by string theory did not come about because it is
mathematically nonsensical. As an infinite component pointlike QFT which
contains operators which communicate between the different floors of an
infinite particle/spin tower it is well-defined and the problem that there
exists a finite number of such objects in 10 spacetime dimensions related to
each other only becomes problematic if it used as the starting point for
making claims about the structure of the universe. The point where the
conceptual confusion starts is that in order to introduce interactions one
uses pictures \textit{as if the pointlike localized infinite component
field\footnote{The attempts to construct infinite component irreducible (in
the described sense) pointlike fields based on higher noncompact group
representations (similar to the O(4,2) hydrogen spectrum) are described in
\cite{Tod}. Unfortunately there was no communication between the two groups of
which only the string construction was successful (a success certainly not
appreciated by string theorists).} would be stringlike,} replacing the lines
in Feynman graphs by the tubelike world-heets traced out by closed strings.
Therefore the recognition that the localization is pointlike does not put an
end to the confusion but rather creates new problems. It is important to note
that, different from Feynman rules, these tube (worldsheet) rules, despite an
intense search by the creme of string theorists over many decades, did not
permit a presentation in terms of operators and states. Having bungled the
localization properties of infinite component fields, at the latest this lack
of presentability of would be hifger orders in terms of operators and states
should have set off alarm bells.

The string theoretic way of metaphoric thinking can perfectly exist outside
string theory. One look at the recent paper \cite{Verlinde} supports this
point; here a string theoretician sets out to colonize territories outside
string theory by treating them with similar Zeitgeist-compatible metaphors.

Perhaps the most spectacular episode triggered by string theory is the fray
which developed around the anti-De Sitter--conformal field theory (AdS-CFT)
correspondence, an issue which, although not directly related to string
theory, suddenly obtained prominence as its alledged consequence. Within a
short time string theorists managed to convert this subject into something
mystical, if not to say surreal.

The subtle point of this correspondence is the radical change of the spacetime
localization involved in the \textit{spacetime reordering of quantum matter}
passing from AdS spacetime to a lower dimensional CFT. Since physics is not
only determined by the abstract quantum matter (e.g. CCR or CAR or any other
matter characterized by its abstract spacetime independent properties), but
also by its spacetime ordering, some physical properties do change with the
spacetime reordering in passing from $AdS_{5}$ to$~CFT_{4}.$ The relevant
question is how much can they change if the abstract matter which is ordered
according to the causal locality in different spacetimes with different
dimensions remains the same? The answer is, that although there is no
correspondence (isomorphism) between pointlike fields, there is one between
tertain operator algebras which are generated by pointlike
fields\footnote{Although there is no theorem that a net of local algebras is
always generated by local fields, the lack of any counterexample suggest that
even if this does not hold for all local nets, it is valid for a large
subset.}. This coarser than pointlike correspondence is sufficient to fix one
side of the correspondence in terms of the other \cite{Rehren}.

The naive expectation about any isomorphismus (correspondence) is that when
one starts from a theory with a \textit{physically acceptable cardinality of
degrees of freedom} (intuitively speaking, one coming from Lagrangian
quantization) and spatially reorders them in such a way that there remains a
local algebraic isomorphism for certain regions\footnote{Neither in the case
of the AdS-CFT correspondence, nor in the case of holographic projections on
the horizon (a nullsurface) of a bulk region, the dimension-changing
holographic map can be expressed in terms of pointlike fields.}, then there
will be \textit{too many degrees of freedom} in case that the reordering leads
to a spacetime of lower dimension as in the AdS$_{5}$--%
$>$%
CFT$_{4}~$correspondence. Although perfectly consistent from a mathematical
viewpoint, this causes serious physical pathologies (Hagedorn temperature or
no thermal states at all, anomalies in the causal propagation etc.)
\cite{Swie}. In the opposite direction CFT$_{4}$--%
$>$%
AdS$_{5}$ the resulting AdS theory obtained from a physical CFT model will be
too "anemic" concerning its degrees of freedom in order to be of any direct
physical interest (the degrees of freedom hover near the boundary). This is
the content of a rigorous mathematical theorem \cite{Rehren} and can be
explicitly illustrated in terms of a free field AdS model \cite{Du-Re}. For
example the CFT theory one obtains as an image under the correspondence from a
free massive AdS model is a generalized free CFT theory with an increasing
Kall\'{e}n-Lehmann spectral function (a power law, depending on the AdS mass)
which violates the causal shadow property and has no physical thermal states.
Far from being a disease of this particular model, it is a structural property
of the correspondence itself.

The Maldacena conjecture \cite{Mal} is more specifically places a concrete
supersymmetric Yang-Mills theory\footnote{If the supersymmetric N=4 Yang-Mills
theory would be the only 4-dimensional CFT, then the correspondence would be
unique.} on the CFT side of the correspondence and expects a supersymmetric
gravity model on the AdS side (suggested by string theory). The prerequiste
for conformal invariance is the vanishing of the beta function. Rigorous
proofs for the absence of radiative corrections and in particular the
vanishing of the Beta functions in certain models were given in the 70s by
combining Callen-Symanzik equations with Ward identities \cite{Lo1}\cite{Lo2}.
Apparently the knowledge about these techniques has been lost, the new order
by order or lightcone quantization attempts applied to the supsersymmetric N=4
Yang Mills model are unconvincing.

But our main criticism is independent of these weaknesses and concerns the
phase space degree of freedom issue which stands in contradiction to the
underlying tacit assumption that both sides represent physical theories. The
above theorem says that this is structurally impossible; if one side is
physical, the other is a purely mathematical chimera which however still may
be useful in order to study certain physical properties of the physical side
which in the original description were not easily accessible.

Since the Maldacena statement is only a conjecture as compared to Rehren's
theorem, there is no paradox here. What renders the whole situation delicate
from a sociological viewpoint however is the fact that meanwhile more than
6000 papers have been written in support of Maldacena's conjecture (but, as
expected, without any conclusion about its validity) and the saying that so
many people cannot err is, as well-known, one of the most accepted
vernaculars. It is hard to think of a more convincing illustration about the
loss of solid scientific knowledge than this episode around the Maldacena correspondence.

The discovery that instead of the finite phase space degrees of freedom in QM
(one per unit phase space cell of size $\hbar$), the cardinality of degrees of
freedom in QFT is different, namely "mildly infinite" (compact, nuclear) was
made in the 60s \cite{Ha-Sw}\cite{Swie}. In the spirit of this article it is
important to emphasize that this difference is a consequence of the different
concepts of localization \cite{interface}. If one compresses the O(4,2)
symmetric degrees of freedom from a physical density in a five-dimensional
spacetime into four dimensions, then there are "too many phase space degrees"
in order to sustain the causal propagation property which is the LQP version
of the classical causal Cauchy propagation. With too many phase space degrees
of freedom the quantum causal shadow property $\mathcal{A(O})=\mathcal{A(O}%
^{\prime\prime}),$ where $\mathcal{O}^{\prime\prime}$ is the causal completion
of the spacetime region $\mathcal{O}$ (the causal complement taken twice), is
being violated; the right hand side is bigger.

From the viewpoint of somebody whose intuitive understanding of QFT comes from
Lagrangian quantization which formally obey this property, the violation may
appears mysterious. The only way he can uphold his picture of propagation is
by using a metaphor that some degrees of freedom enter "sideways" from an
extra dimension or from another universe ("poltergeist degrees of freedom").
Within the present Zeitgeist inspired by string theory, where metaphoric
arguments are en vogue and extra dimensions and multiverses are concepts on
which articles are written, this only sounds like a harmless addition. 

The problem is that deep concepts as the cardinality of degrees of freedom
\cite{Ha-Sw}\cite{Bu-Wi} and their preservation in correspondences between QFT
in different spacetimes have vanished from the conceptual screen of the 80s so
that especially those who work on holographic problems are not aware of their
existence. The notion that metaphoric arguments should at most be tolerated as
placeholders in a conceptual emergency for a limited time has been lost. A
more detailed recent presentation of this phase space degrees of freedom issue
can be found in \cite{Swie}.

Part of the problem of holographic spacetime reordering of quantum matter is
that it is too radical in order to allow a formulation in terms of the
standard setting of QFT using individual pointlike fields; there is however no
problem to express this in terms of operator algebras associated with suitable
causally closed regions \cite{Rehren}.

The only kind of holography which complies with the thinning out of phase
space degrees of freedom is the holography onto nullsurfaces i.e. the
holographic projection of bulk QFT onto causal or event horizons. In that case
the reduction of degrees of freedom goes hand in hand with a reduction of
symmetry: the symmetry of a lightfront is a 7-parameter subgroup of the
Poincar\'{e} group and the problem of "filling up" the degrees of freedom to
their orginal strength is equivalent to knowing the action of the remaining
Poincar\'{e} transformations on the lightfront degrees of freedom.
Equivalently it would suffice to know the lightfront theory in a "GPS manner"
in different positions; in d=1+3 not more than three different positions are
necessary \cite{interface}.

The problems which led to a derailment of a large part of particle theory can
however not fully explain why the comperativly healthy standard model, after
impressive initial gains, entered a period of stagnation. For almost 4 decades
there has been not a single conceptual addition to the age-old central
problems of gluon and quark confinement and the Schwinger-Higgs screening
mechanism. Such a situation is certainly unique in the more then 8 decades
lasting history of particle physics. In some cases there was even a regress in
that earlier promising ideas have been lost in the maelstrom of time
\cite{Swie}.

If there was any influence of S-matrix approach on the standard model
research, it certainly was not of a hepful kind. Rather the perilous charm,
which a TOE supported by prominent community members, exerts on intelligent
and zealous newcomers could have been one reason why the standard model
research may not have attracted the brightest minds; not to mention the
considerable material support enjoyed by string-related research; a closely
related argument is the prediction of the leading string theorists that the
standard model has to appear anyhow as a "low energy effective theory" of a
TOE. Finally there is a widespread but misleading opinion that the remaining
theoretical problems of the standard model are basically of a computational
nature; this is strengthened by the credo that QFT is a reasonably well
understood low energy footnote of string theory.

These beliefs have eroded the enthusiasm for new conceptual investments. A
serious obstacle against a conceptual renewal is the fact that the teaching of
QFT has fallen back behind what can be found in books written before 1980 e.g.
in the book of Itzykson and Zuber. More recent books often appear as a kind of
QFT filtered through string theory glasses. It is nearly impossible to start
research on important conceptual problems (as the problem of the crossing
property in this paper) on the basis of contemporary books on QFT. This has
led to a situation in which the number of people who know QFT sufficiently
well in order to contribute to a conceptual progress of QFT has shrunk to a
few individuals in an advanced age.

Speculative proposals with little conceptual support but a lot of public
attraction were of course made at all times; particle theory by its very
nature is a highly speculative science where it is necessary (at least once in
a while) to take a dive into the "blue yonder". What was however different
during the last 4 decades of dominance of string theory, is that the critical
counterweight, which had quite a tradition in the Streitkultur of the old
continent, was not available when it was most needed. The leading figures in
mathematical physics and (algebraic) quantum field theory who are in the
possession of the necessary conceptual insight to play this indispensible
critical role did not enter the fray.

At the beginning the phenomenonological proposals (the Regge trajectory
setting) were far removed from any structure which one could relate with known
principles of relativistic quantum theory, and when the sudden transition to a
pretended fundamental TOE took place\footnote{The begin of modern string
theory has a date, it is the week in Paris in 1974 when Scherk and Schwarz
\cite{S-S} wrote up their famous paper. Underlining the rapidity of change one
may call it the Bartholomew-like massacre of the old string theory which
started with phenomenology of Regge trajectories.}, the uncritical spread of
the new string theory was too rapid, so that there was hardly time for a
critical discourse. The string protagonists occupied research and university
positions within a short time, and often their only credentials were that they
are working on the most important millennium theory.

There are of course others who understand more or less the causes behind the
derailment. In some of their articles one even finds the statements that
strings are, contrary to their name, really point-localized objects. But since
no critical conclusions are drawn; such articles do not create frictions with
their string theory colleagues. They are tolerated, even when they contribute
jointly to the same book \cite{FRS}), as the kind of critical remarks which
show that string theory is a living science. As long as they do not lead to a
serious conceptual encounter whose outcome could threaten the continued
existence of a more than 40 years lasting development in particle physics, the
present stalemate will continue. The fruitful Streitkultur belonged to the
bygone "golden age" of critical engagement in particle theory.

Similar arguments apply to the sociological \cite{Woit}\cite{Smolin} and
philosophical \cite{He} critique of string theory. Whereas scientific critique
may have the power to erode metaphoric constructs, sociological and
philosophical arguments do not constitute any danger to the popularity of
string theory and certainly have nothing in common with a critical engagement
within a scientific Streitkultur; to the contrary they lead to a profitable
symbiosis between string propagandists and their critics, with the latter
running the risk of loosing their subject without the presence of the former.

Reading the books and articles of the aforementioned authors, the following
questions comes to one's mind. Why can't a theory which has strong conceptual
credentials be explored for whatever time is necessary to get to its limits,
and isn't a consistent theory which, as claimed by string theorists,
incorporates the existing one as a limiting case an interesting goal even if
it does not describe reality? And is observational agreement the only
criterion for evaluating a new theory? The old (pre-oxigen) phlogiston theory
of burning which dominated for many decades shows that a wrong theory may be
able to live for a long time in reasonable agreement with observational facts,
especially if it explains sufficienty many observed phenomena. \textit{The
only kind of critique which a theoretician must take serious in the long run
is one which, as presented in this paper, demonstrates that a theory is
conceptually flawed.}

All these observations show that the adventurous journey that started more
than 4 decades ago with some misunderstandings in the particle-field relation
around the crossing property, has grown into a profound crisis of particle
physics. The resulting metaphoric discourse of placing superficial conclusions
based on calculations done outside any conceptual control above profound
critical evaluations is not any more confined to ST; the concomittant
sociological phenomenon around the AdS-CFT issue is a clear indication of the
spread of the crisis beyond the borders of string theory.

The disappearance of criticism has led to a new culture of establishing a
scientific truth starting from a conjecture and ending after several
reformulations and turns with the acceptance within a community at the level
of a theorem. This process has been insightfully described in a series of
essays by a young string theorist \cite{Zapata}. The author, Oswaldo Zapata,
has an ambivalent position with respect to string theory; having been raised
with string theory and being aware about his limitations with respect to QFT,
he knows that he cannot confront it on its \textit{scientific} truth content.
Instead he carefully analyzes the sociological aspects of its discourse and
comes to remarkable conclusions. His aim is to understand how his fellow
string theorists, having disposed of classical methods of establishing
theorems, arrive at what they consider as truths, and how they present their
results without becoming subjectively dishonest within the community and to
the outside world. He does this by studying changes in the string communities
discourse from conjectures to theorems during a time in which there was no
change in the facts.

Interestingly enough he gives the strongest argument for his thesis about the
relation of the string community to facts involuntary by not referring to the
aforementioned rigorous theorems about AdS-CFT. They are all in the public
domain, but their conceptual mathematical content \cite{Rehren} is not known
by the community members because most of them are not on a level on which they
can understand structural theorems on local quantum physics. This shows that
the control of the community over facts does not end at what is coming from
the inside (which Zapata as an insider of this community is well aware of),
but it extends also to shielding inconvenient theorems from the outside in the
most possible honest manner, namely by ignorance about large parts of QFT. In
this way even Zapata remains uninformed that his critical sociological
observations about the discourse of the string community have a profound
scientific counterpart.

Reading Zapata's essay may not help to learn about conceptual errors of string
theory. But his method is very successful in exposing the surreal aspect which
accompanies the string community's almost messianic "end of the millennium
belief" in a TOE. His account of how a metaphoric conjecture ends after
several sweeps through the community as a community-accepted fact is truely
remarkable. It shows that some individuals of the string generation, having
been deprived of a critical conceptual scientific basis, can still make
fascinating critical observations about the logic and sociology of the
discourse within the string community.

It is quite revealing that Zapata takes a dim view on some missing arguments
in two books by Lee Smolin and Peter Woit \cite{Smolin}\cite{Woit}. These
authors take a critical look at the dominant position of string theory and
explain very well the sociological reasons why younger people uncritically
internalize the catechism of string theory. \textit{But they never explain why
respectable older people, who are under no such career pressures (especially
those who are the main string proselytizers mentioned before) believe in the
validity of the theory.} It is of course common practice to blame the
foot-soldiers (in the present context, the young partisans of string theory)
and the propaganda division (Brian Green and others), but spare the generals;
there should be no place for this attitude in particle physics.

It would be wishful thinking that articles as the present one or the essay of
Zapata could have an influence on the tide of events. But they provide a
valuable help for historians and philosophers of science to analyze what went
on in particle theory during a substantial part of the 20$^{th}$ and the
beginning of the 21st century.

Since readers need some encouragement in the conclusions, the present essay
should not end in a downbeat mood. There are some interesting new developments
around higher spin field, in particular massless fields. They start from the
observation made in the appendix in (\ref{line}) where it was mentioned that
the reduced possibilities for $(m=0,s)$ with $s=1,2$ which exclude covariant
vector potentials and $g_{\mu\nu}$ tensors, can be complemented to the full
spinorial formalism (so that the massless situation is on par with massive
case) if one permits semiinfinite string localization \cite{charge}. This
leads to a new way of looking at the problems behind gauge theory. Already in
the abelian case of QED for which it has been known for a long time that
electrically charged states are semiinfinite string-localized (associated to
infraparticles), the new setting incorporates the perturbative aspects of
these physical charge-carrying fields into the formalism i.e. they do not have
to be defined by hand outside the perturbation formalism as in the famous
stringlike formulas of Dirac-Jordan-Mandelstam \cite{Swie}\cite{Jor}. The
Higgs model results as a Schwinger-Higgs screening of the electric charge of a
scalar fields and leads to a theory in which the massive matter field is
neutral (real) and pointlike localized \cite{Swie}. The new conceptual frame
of modular localization promises to lead to a significant enlargement of the
range of renormalizability \cite{charge}.

\section{Appendix: a sketch of modular localization}

\subsection{Modular localization of states}

The simplest context for a presentation of the idea of modular localization is
the Wigner representation theory of the Poincar\'{e} group. It has been
realized by Brunetti, Guido and Longo \cite{BGL} \footnote{With somewhat
different motivations and lesser mathematical rigor see also \cite{Sch}.}
there is a natural localization structure on the Wigner representation space
for any positive energy representation of the proper Poincar\'{e} group. Upon
second quantization this representation theoretically determined localization
theory gives rise to a local net of operator algebras on the Wigner-Fock space
over the Wigner representation space.

The starting point is an irreducible representation $U_{1}~$of the
Poincar\'{e} group on a Hilbert space $H_{1}$ that after "second quantization"
becomes the single-particle subspace of the Hilbert space (Wigner-Fock-space)
$H_{WF}$ of the field\footnote{The construction works for arbitrary positive
energy representations, not only irreducible ones.}. The construction proceeds
according to the following steps \cite{BGL}\cite{Fa-Sc}\cite{MSY}. To maintain
simplicity, we limit our presentation to the spinless bosonic situation.

One first fixes a reference wedge region, e.g. $W_{0}=\{x\in\mathbb{R}%
^{d},x^{d-1}>\left\vert x^{0}\right\vert \}$ and considers the one-parametric
L-boost group (the hyperbolic rotation by $\chi$ in the $x^{d-1}-x^{0}$ plane)
which leaves $W_{0}$ invariant; one also needs the reflection $j_{W_{0}}$
across the edge of the wedge which is apart from a $\pi$-rotation in the
transverse plane identical to the TCP transformation. The Wigner
representation is then used to define two commuting wedge-affiliated
operators
\begin{equation}
\mathfrak{\delta}_{W_{0}}^{it}=\mathfrak{u}(0,\Lambda_{W_{0}}(\chi=-2\pi
t)),~\mathfrak{j}_{W_{0}}=\mathfrak{u}(0,j_{W_{0}})
\end{equation}
where attention should be paid to the fact that in a positive energy
representation any operator which inverts time is necessarily
antilinear\footnote{The wedge reflection $\mathfrak{j}_{W_{0}}$ differs from
the TCP operator only by a $\pi$-rotation around the W$_{0}$ axis.}. A unitary
one- parametric strongly continuous subgroup as $\delta_{W_{0}}^{it}$ can be
written in terms of a selfadjoint generator as $\delta_{W_{0}}^{it}%
=e^{-itK_{W_{0}}}$ and therefore permits an "analytic continuation" in $t$ to
an unbounded densely defined positive operators $\delta_{W_{0}}^{s}$.
Poincar\'{e} covariance allows to extend these definitions to wedges in
general position, and intersections of wedges lead to the definitions for
general localization regions (see later). Since the localization is clear from
the context, a generic notation without subscripts will be used. With the help
of this operator one defines the unbounded antilinear operator $\mathfrak{s}$
which has the same dense domain.%
\begin{align}
&  \mathfrak{s}=\mathfrak{j\delta}^{\frac{1}{2}},~dom\mathfrak{s}%
=dom\mathfrak{\delta}^{\frac{1}{2}}\\
&  \mathfrak{j\delta}^{\frac{1}{2}}\mathfrak{j}\mathfrak{=\delta}^{-\frac
{1}{2}}%
\end{align}

Whereas the unitary operator $\delta^{it}$ commutes with the reflection, the
antiunitarity of the reflection causes a change of sign in the analytic
continuation as written in the second line. This leads to the involutivity of
the s-operator as well as the identity of its range with its domain
\begin{align*}
\mathfrak{s}^{2}  &  \subset\mathbf{1}\\
dom~\mathfrak{s}  &  =ran~\mathfrak{s}%
\end{align*}
$\mathbf{.}$ Such operators which are unbounded and yet involutive on their
domain are quite unusual; according to my best knowledge they only appear in
modular theory and it is precisely these unusual aspects which are capable to
encode geometric localization properties into domain properties of abstract
quantum operators. The more general algebraic context in which Tomita
discovered modular theory will be mentioned later.

The idempotency means that the s-operator has $\pm1$ eigenspaces; since it is
antilinear the +space multiplied with $i$ changes the sign and becomes the -
space; hence it suffices to introduce a notation for just one of the two
eigenspaces%
\begin{align}
\mathfrak{K}(W)  &  =\{domain~of~\Delta_{W}^{\frac{1}{2}},~\mathfrak{s}%
_{W}\psi=\psi\}\\
\mathfrak{j}_{W}\mathfrak{K}(W)  &  =\mathfrak{K}(W^{\prime})=\mathfrak{K}%
(W)^{\prime},\text{ }duality\nonumber\\
\overline{\mathfrak{K}(W)+i\mathfrak{K}(W)}  &  =H_{1},\text{ }\mathfrak{K}%
(W)\cap i\mathfrak{K}(W)=0\nonumber
\end{align}

It is important to be aware that, unlike QM, we are dealing here with real
(closed) subspaces $\mathfrak{K}$ of the complex one-particle Wigner
representation space $H_{1}$.

An alternative which avoids the use of real subspaces is to directly work with
complex dense subspaces as in the third line. Introducing the graph norm of
the dense space, the complex subspace in the third line becomes a Hilbert
space in its own right. The upper dash on regions in the second line denotes
the causal disjoint (which is the opposite wedge) whereas the dash on real
subspaces means the simplectic complement with respect to the simplectic form
$Im(\cdot,\cdot)$ on $H_{1}.$

The two equations in the third line are the defining property of what is
called the \textit{standardness} of a subspace\footnote{According to the
Reeh-Schlieder theorem a local algebra $\mathcal{A(O})$ in QFT is in standard
position with respect to the vacuum i.e. it acts on the vacuum in a cyclic and
separating manner. The spatial standardness, which follows directly from
Wigner representation theory, is just the one-particle projection of the
Reeh-Schlieder property.}; any standard K-space permits to define an abstract
s-operator as follows%
\begin{align}
\mathfrak{s}(\psi+i\varphi)  &  =\psi-i\varphi\\
\mathfrak{s}  &  =\mathfrak{j}\delta^{\frac{1}{2}}\nonumber
\end{align}
whose polar decomposition (written in the second line) returns the two modular
objects $\delta^{it}$ and $\mathfrak{j}$ which outside the context of the
Poincar\'{e} group has in general no geometric significance. The domain of the
Tomita $s$-operator is the same as the domain of $\delta^{\frac{1}{2}}$ namely
the real sum of the K space and its imaginary multiple. Note that in the
present context this domain is determined solely by Wigner's group
representation theory.

It is easy to obtain a net of K-spaces by $U(a,\Lambda)$-transforming the
K-space for the distinguished $W_{0}.$ A bit more tricky is the construction
of sharper localized subspaces via intersections
\begin{equation}
\mathfrak{K}(\mathcal{O})=%
{\displaystyle\bigcap\limits_{W\supset\mathcal{O}}}
\mathfrak{K}(W)
\end{equation}
where $\mathcal{O}$ denotes a causally complete smaller region (noncompact
spacelike cone, compact double cone). Intersection may not be standard, in
fact they may be zero in which case the theory allows localization in $W$ (it
always does) but not in $\mathcal{O}.$ Such a theory is still causal but not
local in the sense that its associated free fields are pointlike.

There are three classes of irreducible positive energy representation, the
family of massive representations $(m>0,s)$ with half-integer spin $s$ and the
family of massless representation which consists of two subfamilies with quite
different properties namely the $(0,h),$ $h$ half-integer class (the neutrino,
photon class), and the rather large class of $(0,\kappa>0)$ infinite helicity
representations parametrized by a continuous-valued Casimir invariant $\kappa$
\cite{MSY}$.$

For the first two classes the $\mathfrak{K}$-space is standard for arbitrarily
small $\mathcal{O}$, but this is definitely not the case for the infinite
helicity family for which the compact localization spaces turn out to be
trivial\footnote{It is quite easy to prove the standardness for spacelike cone
localization (leading to singular stringlike generating fields) just from the
positive energy property which is shared by all three families \cite{BGL}.}.
Their tightest localization, which still permits nontrivial (in fact standard)
$\mathfrak{K}$-spaces for \textit{all} positive energy representations, is
that of a \textit{spacelike cone }\cite{BGL} with an arbitrary small opening
angle whose core is a \textit{semiinfinite string} \cite{MSY}; after "second
quantization" (see next subsection) these strings become the localization
region of string-like localized covariant generating fields\footnote{The
epithet "generating" refers to the tightest localized singular field
(operator-valued distribution) which generates the spacetime-indexed net of
algebras in a QFT. In the case of localization of states the generators are
state-valued distributions.}. The modular localization of states, which is
governed by the unitary representation theory of the Poincar\'{e} group, has
only two kind of generators: pointlike state and semiinfinite stringlike
states; generating states of higher dimensionality ("brane states") are not needed.

Although the observation that the third Wigner representation class is not
pointlike generated was made many decades ago, the statement that it is
semiinfinite string-generated and that this is the worst possible case of
state localization is of a more recent vintage \cite{BGL} since it needs the
application of the modular theory.

There is a very subtle aspect of modular localization which one encounters in
the second Wigner representation class of \textit{massless finite helicity
representations} (the photon, graviton..class). Whereas in the massive case
all spinorial fields $\Psi^{(A,\dot{B})}$ the relation of the physical spin
$s$ with the two spinorial indices follows the naive angular momentum
composition rules \cite{Weinberg}\cite{charge}%
\begin{align}
\left\vert A-\dot{B}\right\vert  &  \leq s\leq\left\vert A+\dot{B}\right\vert
,\text{ }m>0\label{line}\\
s  &  =\left\vert A-\dot{B}\right\vert ,~m=0\nonumber
\end{align}
the second line contains the considerably reduced but still infinite number of
spinorial descriptions for zero mass and finite helicity \cite{MSY}.

By using the recourse of string-localized generators $\Psi^{(A,\dot{B})}(x,e)$
one can \textit{restore the full spinorial spectrum} for a given $s$ i. e. one
can move from the second line to the first line in (\ref{line}) by relaxing
the localization. Even in the massive situation where pointlike generators
exist but have short distance singularities which increase with spin. there
may be good reasons (lowering of short distance dimension down to sdd=1) to
use string-like generators. In all cases these generators are covariant and "string-local"%

\begin{align}
U(\Lambda)\Psi^{(A,\dot{B})}(x,e)U(\Lambda)  &  =D^{(A,\dot{B})}(\Lambda
^{-1})\Psi^{(A,\dot{B})}(\Lambda x,\Lambda e)\\
\left[  \Psi^{(A,\dot{B})}(x,e),\Psi^{(A^{\prime},\dot{B}^{\prime})}%
(x^{\prime},e^{\prime}\right]  _{\pm}  &  =0,~x+\mathbb{R}_{+}e><x^{\prime
}+\mathbb{R}_{+}e^{\prime}\nonumber
\end{align}
Here the unit vector $e$ is the spacelike direction of the semiinfinite string
and the last line expresses the spacelike fermionic/bosonic spacelike
commutation. The best known illustration is the ($m=0,s=1$) representation; in
this case it is well-known that although a generating pointlike field strength
exists, there is no pointlike vectorpotential. The modular localization
approach offers as a substitute a stringlike vector potential $A_{\mu}(x,e).$
In the case ($m=0,s=2$) the "field strength" is a fourth degree tensor which
has the symmetry properties of the Riemann tensor; in fact it is often
referred to as the linearized Riemann tensor. In this case the
string-localized potential is of the form $g_{\mu\nu}(x,e)$ i.e. resembles the
metric tensor of general relativity. The consequences of this localization for
a reformulation of gauge theory will be taken up in a separate subsection.

The most radical form of string localization occurs in the massless infinite
spin representation family. In that case the representation space does not
contain any pointlike localized generators which play the role of field
strength, hence such a theory is without any local observables.

A different kind of spacelike string-localization arises in d=1+2 Wigner
representations with anomalous spin \cite{Mu1}. The amazing power of this
modular localization approach is that it preempts the spin-statistics
connection already in the one-particle setting, namely if s is the spin of the
particle (which in d=1+2 may take on any real value) then one finds for the
connection of the simplectic complement with the causal complement the
generalized duality relation
\[
\mathfrak{K}(\mathcal{O}^{\prime})=Z\mathfrak{K}(\mathcal{O})^{\prime}%
\]
where the square of the twist operator $Z=e^{\pi is}~$is easily seen (by the
connection of Wigner representation theory with the two-point function) to
lead to the statistics phase: $Z^{2}=$ statistics phase \cite{Mu1}. The
one-particle modular theory also leads to a relation which may be considered
as the proto-form of crossing in the one-particle space%
\begin{equation}
\mathfrak{\Delta}^{\frac{1}{2}}\mathfrak{j\psi(}p\mathfrak{)=}\overline
{\mathfrak{\psi(-}p\mathfrak{)}}%
\end{equation}
in words the $\mathfrak{\Delta}^{\frac{1}{2}}\mathfrak{j=s}^{\ast}%
~$transformed wave function is equal to the complex conjugate (antiparticle)
and from forward to backward mass shell analytically continued (through the
connecting complex mass shell) wave function.

That one never has to go beyond string localized wave functions (and in fact,
apart from those mentioned cases, even never beyond point localization) in
order to obtain the generating fields for a QFT is remarkable in view of the
many attempts to introduce extended objects into QFT.

It should be clear that modular localization, which is formulated in terms of
either real or dense complex subspaces, cannot be connected with probabilities
and projectors. It is rather related to causal localization aspects and the
standardness of the K-space for a compact region is nothing else then the
one-particle version of the Reeh-Schlieder property. Fortunately one needs the
probability and the projectors from the BNW localization only for asymptotic
timelike scattering distances in which case they become frame-independent and
the discrepancy with modular localization disappears.

\subsection{Localized subalgebras}

A net of real subspaces $\mathfrak{K}(\mathcal{O})$ $\subset$ $H_{1}$ for an
finite spin (helicity) Wigner representation can be "second
quantized"\footnote{The terminology 2$^{nd}$ quantization is a misdemeanor
since one is dealing with a rigorously defined functor within QT which has
little in common with the artful use of that parallellism to classical theory
called "quantization". In Edward Nelson's words: (first) quantization is a
mystery, but second quantization is a functor.} via the CCR (Weyl)
respectively CAR quantization functor; in this way one obtains a covariant
$\mathcal{O}$-indexed net of von Neumann algebras $\mathcal{A(O)}$ acting on
the bosonic or fermionic Fock space $H=Fock(H_{1})$ built over the
one-particle Wigner space $H_{1}.$ For integer spin/helicity values the
modular localization in Wigner space implies the identification of the
simplectic complement with the geometric complement in the sense of
relativistic causality, i.e. $\mathfrak{K}(\mathcal{O})^{\prime}%
=\mathfrak{K}(\mathcal{O}^{\prime})$ (spatial Haag duality in $H_{1}$). The
Weyl functor takes the spatial version of Haag duality into its algebraic
counterpart. One proceeds as follows: for each Wigner wave function
$\varphi\in H_{1}$ the associated (unitary) Weyl operator is defined as%
\begin{align}
Weyl(\varphi)  &  :=expi\{a^{\ast}(\varphi)+a(\varphi)\},Weyl(\varphi)\in
B(H)\\
\mathcal{A(O})  &  :=alg\{Weyl(\varphi)|\varphi\in\mathfrak{K}(\mathcal{O}%
)\}^{^{\prime\prime}},~~\mathcal{A(O})^{\prime}=\mathcal{A(O}^{\prime
})\nonumber
\end{align}
where $a^{\#}(\varphi)$ are the usual Fock space creation and annihilation
operators of a Wigner particle in the wave function $\varphi$. We then define
the von Neumann algebra corresponding to the localization region $\mathcal{O}$
in terms of the operator algebra generated by the functorial image of the
modular constructed localized subspace $\mathfrak{K}(\mathcal{O})$ as written
in the second line. By the von Neumann double commutant theorem, our generated
operator algebra is weakly closed by definition.

The functorial relation between real subspaces and von Neumann algebras via
the Weyl functor preserves the causal localization structure, hence the
spatial duality passes to its algebraic counterpart. The functor also commutes
with the process of sharpening localization through intersections $\cap$
according to $K(\mathcal{O})=\cap_{W\supset O}K(W),~\mathcal{A(O}%
)=\cap_{W\supset O}\mathcal{A}(W)$ as expressed in the commuting diagram%
\begin{align}
&  \left\{  K(W)\right\}  _{W}\longrightarrow\left\{  \mathcal{A}(W)\right\}
_{W}\\
&  \ \ \downarrow\cap~~~\ \ \ \ \ \ \ \ \ \ ~\ ~\downarrow\cap\nonumber\\
~~  &  \ \ \ K(\mathcal{O})\ \ \ \longrightarrow\ \ ~\mathcal{A(O})\nonumber
\end{align}
Here the vertical arrows denote the tightening of localization by
intersection, whereas the horizontal ones denote the action of the Weyl functor.

The case of half-integer spin representations is analogous \cite{Fa-Sc}, apart
from the fact that there is a mismatch between the causal and simplectic
complements to be taken care of by a \textit{twist operator} $\mathcal{Z}$ and
as a result one arrives at the CAR functor instead of the Weyl functor.

In case of the large family of irreducible zero mass infinite spin
representations for which the lightlike little group, different from the
finite helicity representations, is faithfully represented, the finitely
localized K-spaces are trivial $\mathfrak{K}(\mathcal{O})=\{0\}$ and the
\textit{most tightly localized nontrivial spaces} \textit{are of the form}
$\mathfrak{K}(\mathcal{C})$ for $\mathcal{C}$ a \textit{spacelike cone}. As a
double cone contracts to its pointlike core, the core of a spacelike cone
$\mathcal{C}$ is a \textit{covariant spacelike semiinfinite string}. The above
functorial construction works the same way for the Wigner infinite spin
representation, except that there are no nontrivial compactly localized
algebras with a smaller localization than $\mathcal{A(C})$, and there are no
generating fields which are sharper localized than a semiinfinite spacelike
string. Point- (or string-) like covariant fields are singular generators of
these algebras i.e. operator-valued distributions. Stringlike generators,
which are also available in the pointlike case, turn out to have an improved
short distance behavior; whereas e.g. the short distance dimension of a free
pointlike vectorfield is $sddA_{\mu}(x)=2,$ its stringlike counterpart has
$sddA_{\mu}(x,e)=1~$\cite{MSY} thanks to the fact that the vacuum fluctuations
are spread into $e$ as well. Covariant representations are constructed from
the unique Wigner representation by so called intertwiners between the
canonical and the many possible covariant (dotted-undotted spinorial
representations of the L-group) representations. Whereas for pointlike
generators this is done by group theoretic methods as in \cite{Weinberg}, the
construction of string-like intertwiners require the use of modular
localization \cite{MSY}. The Euler-Lagrange formalism plays no role in these
construction since the causal aspect of hyperbolic differential propagation
are fully taken care of by modular localization.

A basis of local covariant field coordinatizations is defined by Wick
composites of the free fields. The string-like fields do not follow the
classical behavior; already before introducing composites one has a continuous
family of non-classical intertwiners between the unique Wigner infinite spin
representation and the \textit{continuously many covariant string
interwiners}. These non-classical aspects, in particular the absence of a
Lagrangian, are the reason why their spacetime description in terms of
semiinfinite string fields has been discovered only recently and not at the
time of Jordan's field quantization nor at the time of Wigner's representation theory.

Using the standard notation $\Gamma$ for the second quantization functor which
maps real localized (one-particle) subspaces into localized von Neumann
algebras, and extending this functor in a natural way to include the
functorial images of the $\mathfrak{K}(\mathcal{O})$-associated objects
$s,\delta,j$ (denoted by $S,\Delta,J),$ one arrives at the Tomita Takesaki
theory of the interaction-free local algebra ($\mathcal{A(O}),\Omega$) in
standard position\footnote{The functor $\Gamma$ preserves the standardness
i.e. maps the spatial one-particle standardness into its algebraic
counterpart.}%
\begin{align}
&  H_{Fock}=\Gamma(H_{1})=e^{H_{1}},~\left(  e^{h},e^{k}\right)
=e^{(h,k)}\label{mod}\\
&  \Delta=\Gamma(\delta),~J=\Gamma(j),~S=\Gamma(s)\nonumber\\
&  SA\Omega=A^{\ast}\Omega,~A\in\mathcal{A}(O),~S=J\Delta^{\frac{1}{2}%
}\nonumber
\end{align}

With this result we got to the core statement of the Tomita-Takesaki theorem
which is a statement about the action of the two modular objects $\Delta^{it}$
and $J$ on the algebra%
\begin{align}
\sigma_{t}(\mathcal{A(O}))  &  \equiv\Delta^{it}\mathcal{A(O})\Delta
^{-it}=\mathcal{A(O})\\
J\mathcal{A(O})J  &  =\mathcal{A(O})^{\prime}=\mathcal{A(O}^{\prime})\nonumber
\end{align}
in words: the reflection $J$ maps an algebra (in standard position) into its
von Neumann commutant and the unitary group $\Delta^{it}$ defines an
one-parametric automorphism-group $\sigma_{t}$ of the algebra. In this form
(but without the last statement involving the geometrical causal complement
$\mathcal{O}^{\prime})$ the theorem hold in complete mathematical generality
for standard pairs ($\mathcal{A},\Omega$). The free fields and their Wick
composites are "coordinatizing" singular generators of this $\mathcal{O}%
$-indexed net of algebras in the sense that the smeared fields $A(f)$ with
$suppf\subset\mathcal{O}$ are (unbounded operators) affiliated with
$\mathcal{A(O}).$

In the above second quantization context the origin of the T-T theorem and its
proof is clear: the spatial symplectic disjoint passes via the functorial
operation to the operator algebraic commutant and the spatial one-particle
modular automorphism goes into its algebraic counterpart. The definition of
the Tomita involution $S$ through its action on the dense set of states
(guarantied by the standardness of $\mathcal{A}$) as $SA\Omega=A^{\ast}\Omega$
and the action of the two modular objects $\Delta,J$ (\ref{mod}) is part of
the general setting of the modular Tomita-Takesaki theory; standardness is the
mathematical terminology for the Reeh-Schlieder property \cite{Haag} i.e. the
existence\footnote{In QFT any finite energy vector (which of course includes
the vacuum) has this property as well as any nondegenerated KMS state. In the
mathematical setting it is shown that standard vectors are "$\delta-$dense" in
$H$.} of a vector $\Omega\in H$ with respect to which the algebra acts cyclic
and has no "annihilators" of $\Omega.$ Naturally the proof of the abstract T-T
theorem in the general setting of operator algebras or even in the more
restricted context of interacting QFT is more involved \cite{Haag}.

The important property which renders this formalism useful beyond free fields
as a new constructive tool in the presence of interactions, is that for
$\left(  \mathcal{A}(W),\Omega\right)  ~$ the antiunitary involution $J$
depends on the interaction, whereas $\Delta^{it}$ continues to be uniquely
fixed by the representation of the Poincar\'{e} group i.e. by the particle
content. In fact it has been known for some \cite{Sch} time that $J$ is
related with its free counterpart $J_{0}$ through the scattering matrix%
\begin{equation}
J=J_{0}S_{scat} \label{scat}%
\end{equation}

This modular role of the scattering matrix as a relative modular invariant
between an interacting theory and its free counterpart comes as a surprise. It
is precisely this role which opens the way for an inverse scattering
construction \cite{inverse} and the constructive approach to factorizing
models \cite{Lech1}. Closely related to this observation is the realization
that the wedge region leads to a coexistence of one particle states in
interacting theories (section 6) with modular localization; namely there is a
dense set of wedge-localized one particle states and their multiparticle
in/out extensions in the interacting theory. With other words the wedge region
is the "smallest" region for which PFGs (vacuum \textbf{p}olarization
\textbf{f}ree \textbf{g}enerators) and their multiparticle generalizations are
available. This is the origin of the crossing property as explained in section 5.

For the construction of a QFT it suffices to specify wedge algebra
$\mathcal{A}(W)$ for one particular wedge $W$ as well as the action of the
Poincar\'{e} group on $\mathcal{A}(W)$ which results in a net of wedge
algebras $\left\{  \mathcal{A}(W)\right\}  _{W\in\mathfrak{W}}.$Knowing a
wedge algebra means knowing its position in the global algebra $\mathcal{A}%
(W)\subset B(H);$ in practice this is achieved by describing $\mathcal{A}(W)$
in terms of generators as explained before in the special case of factorizing
models. By taking suitable intersections of wedge algebras one obtains (in
case the double cone intersections are nontrivial) a net of local observables
i.e. a nontrivial local QFT or (if they are trivial) there is no local QFT
associated with the system of wedge algebras. In this way one is able to
separate the existence proof for a local QFT from the harder problem of the
construction of its pointlike fields\footnote{The necessarily singular
pointlike fields are universal generators for algebras of arbitrary (small)
localization.} via their correlation functions or formfactors. Hence the
construction of a QFT may be seen as a generalization of those ideas which
lead to a proof of the crossing property.

An "observable net" is a spacetime-indexed family of operator algebras
consisting of chargeless operators. By definition these operators fulfill
spacelike commutativity and have, as the vacuum, vanishing charge. There
exists a very deep theory which intrinsically constructs all charged sectors
and combines them to a generally quite large "field-algebra" which in a way
defines the maximal extension of the observable algebra; this is the famous
Doplicher-Haag-Roberts (DHR) superselection theory \cite{Haag}. It explains
statistics and inner symmetries in terms of spacetime localization properties
of the observable net\footnote{There is a complication in low-dimensional
theories in which braid group statistics may occur in which case there is no
sharp separation between inner and spacetime symmetries. Nevertheless these
representations appear in the DHR theory \cite{Haag}.}. From a point of view
of principles of QFT one can show that in more than 3 dimensions all compact
groups can appear. What does not appear in this classification is supersymmetry.

A slight reformulation of this algebraic setting which leads to a
(philosophically) quite spectacular new view of the core nature of local
quantum physics. Namely it is possible to encode the entire content of QFT
i.e. the net of local observables \textit{as well} as all its superselected
charge sectors and their interpolating charged fields including the
representation of the Poincar\'{e} group acting on it, into a finite set of
copies of the monad (physically interpreted as $\mathcal{A}(W)s$) carefully
positioned in a joint Hilbert space with the help of modular theory, using
concepts of "modular inclusion" and "modular intersection " within a joint
Hilbert space \cite{interface}. The representation theory of the Poincar\'{e}
group and therefore of spacetime itself arises from the joint action of the
individual modular groups in the form of unitary operators in the shared
Hilbert space. This is as close as one can get to how Leibniz envisaged
reality as emerging from \textit{relations} between monads, the monads (here
copies of the unique hyperfinite Type III$_{1}\operatorname{factor}$ algebra)
themselves being structureless\footnote{In this sense they are like points in
geometry except that monads can be mutually included and intersected.}. \ 

It is an interesting open question whether \ a characterization of a QFT in
terms of positioning of a finite number of monads can be extended to curved
spacetime. The recent successful quantum formulation of the principle of local
covariance \cite{BFV} nourishes some hope that this may be the case.

Acknowledgement: I am indepted to Jens Mund who on several occasions gave me
advice on matters of modular localization.

\end{document}